\documentclass{aa}

\usepackage{graphicx}
\usepackage{amsmath}
\usepackage{hyperref} 
\usepackage{lipsum}
\usepackage{txfonts}

\usepackage{etoolbox}
\makeatletter
\patchcmd\@combinedblfloats{\box\@outputbox}{\unvbox\@outputbox}{}{%
  \errmessage{\noexpand\@combinedblfloats could not be patched}%
}%
\makeatother

\begin{document}

\title{Characterization of low-mass companion HD~142527~B
\thanks{Based on observations made with ESO Telescopes at the Paranal Observatory under program ID 084.C-0952 and 093.C-0526.}}

\author{
V.~Christiaens\inst{1,2,3}
\and S.~Casassus\inst{1,3}
\and O.~Absil\inst{2}\thanks{F.R.S-FNRS Research Associate}
\and S.~Kimeswenger\inst{4,5}
\and C.~A.~Gomez~Gonzalez\inst{2}
\and J.~Girard\inst{6}
\and R.~Ram\'irez\inst{1}
\and O.~Wertz\inst{2,7}
\and A.~Zurlo\inst{3,8,9}
\and Z.~Wahhaj\inst{10}
\and  C.~Flores\inst{11}
\and V.~Salinas\inst{12}
\and A.~Jord\'an\inst{13}
\and D.~Mawet\inst{14,15}
}
\institute{Departamento de Astronom\'ia, Universidad de Chile, Casilla 36-D, Santiago, Chile\\
\email{valchris@das.uchile.cl}
\and
Space sciences, Technologies \& Astrophysics Research (STAR) Institute, Universit\'e de Li\`ege, All\'ee du Six Ao\^ut 19c, B-4000 Sart Tilman, Belgium
\and
Millenium Nucleus "Protoplanetary Disks", Chile
\and
Instituto de Astronom\'ia, Universidad Cat\'olica del Norte, Avenida Angamos 0610, Antofagasta, Chile
\and
Institut f{\"u}r Astro-- und Teilchenpysik, Leopold--Franzens Universit{\"a}t Innsbruck, Technikerstr. 25, A-6020 Innsbruck, Austria
\and
Space Telescope Science Institute, 3700 San Martin Dr. Baltimore, MD 21218, USA
\and
Argelander-Institut f\"ur Astronomie, Universit\"at Bonn, Auf dem H\"ugel 71, D-53121 Bonn, Germany
\and
N\'ucleo de Astronom\'ia, Facultad de Ingenier\'ia y Ciencias, Universidad Diego Portales, Av. Ejercito 441, Santiago, Chile
\and
Escuela de Ingenier\'ia Industrial, Facultad de Ingenier\'ia y Ciencias, Universidad Diego Portales, Av. Ejercito 441, Santiago, Chile
\and
European Southern Observatory, Alonso de C\'ordova 3107, Vitacura, Santiago, Chile
\and
 Institute for Astronomy, University of Hawaii, 640 N. Aohoku Place, Hilo, HI 96720, USA
\and
Department of Physics and Astronomy, Graduate School of Science and Engineering, Kagoshima University, 1-21-35 Korimoto, Kagoshima, Kagoshima 890-0065, Japan
\and
Instituto de Astrof\'isica, Pontificia Universidad Cat\'olica de Chile, Vicu\~na Mackenna 4860, 7820436 Macul, Santiago, Chile
\and
Department of Astronomy, California Institute of Technology, 1200 E. California Blvd, Pasadena, CA 91125, USA
\and
Jet Propulsion Laboratory, 4800 Oak Grove Dr., Pasadena, CA 91109, USA
}

\date{}

\abstract
{The circumstellar disk of the Herbig Fe star HD~142527 is host to several remarkable features 
including a warped inner disk, a 120~au-wide annular gap, a prominent dust trap and several spiral arms.
A low-mass companion, HD~142527~B, was also found orbiting the primary star at $\sim$14~au.} 
{This study aims to better characterize this companion, which could help explain its impact on the peculiar geometry of the disk.} 
{We observed the source with VLT/SINFONI in $H$+$K$ band in pupil-tracking mode. Data were post-processed with several algorithms based on angular differential imaging (ADI).}
{HD~142527~B is conspicuously re-detected in most spectral channels, 
which enables us to extract the first medium-resolution spectrum of a low-mass companion within 0\farcs1~from its central star.
Fitting our spectrum with both template and synthetic spectra suggests that the companion is a young M2.5$\pm$1.0 star with an effective temperature of $3500\pm100$~K, possibly surrounded with a hot (1700~K)  circum-secondary environment.
Pre-main sequence evolutionary tracks provide a mass estimate of $0.34\pm0.06 M_{\sun}$, independent of the presence 
of a hot 
environment. 
However, the estimated stellar radius and age do depend on that assumption; we find a radius of $1.37 \pm 0.05 R_{\sun}$ (resp. $1.96 \pm 0.10 R_{\sun}$) and an age of $1.8^{+1.2}_{-0.5}$~Myr (resp. $0.75 \pm 0.25$~Myr) in the case of the presence (resp. absence) of a hot environment contributing in $H$+$K$.
Our new values for the mass and radius of the companion yield a mass accretion rate of 4.1--5.8 $\times 10^{-9}~M_{\sun}$~yr$^{-1}$ (2--3\% that of the primary).
}
{We have constrained the physical properties of HD~142527~B, thereby illustrating the potential for SINFONI+ADI to characterize faint close-in companions. 
The new spectral type makes HD~142527~B a twin of the well-known TW~Hya T-Tauri star, and the revision of its mass to higher values further supports its role in shaping the disk.
}

\keywords{Protoplanetary disks - Stars: binaries: close - Stars: formation - Stars: low-mass  - Stars: pre-main sequence - Stars: individual: HD~142527}

\maketitle

\section{Introduction}\label{intro}
The advents of the Atacama Large Millimeter Array (ALMA) and extreme adaptive optics high-contrast imaging instruments have recently unveiled a wealth of features in protoplanetary disks, including cavities, rings, spiral arms, warps and asymmetric dust distributions \citep[see][and references therein]{Casassus2016}.
A major challenge now is to connect these observed features to the process of planet formation, which is believed to be concomitant.
Giant planets must indeed form within the first million years of the disk life, before all primordial gas is dissipated \citep{Haisch2001,Hernandez2007a}.
While concentric annular gaps have been observed in protoplanetary disks starting from an early age \citep[e.g.,][]{ALMA2015,Andrews2016}, a fraction of disks also harbor very large gaps or cavities ($>20$ au in radius).
This population overlaps with the class of \emph{transition(al) disks} originally identified from their SEDs \citep[e.g.,][]{Strom1989,Espaillat2007}.
Whether a common origin to the large gaps in those disks can be found is still an open debate.
Interestingly, most of these disks with very large gaps also show strong millimeter flux and significant accretion rates, 
which would hint at an origin related to forming planets rather than, for example,~photo-evaporation \citep[see][and references therein]{Owen2016}.
Nevertheless, this picture still suffers pitfalls, as the standard disk-evolution scenario in the presence of giant planets should lead to a population of millimeter-bright transition disks with no accretion \citep[e.g.,][]{Rosotti2015}, which is not observed.

In this context, the system of \object{HD~142527} constitutes a remarkable case study for protoplanetary disk evolution 
and possible on-going planet formation.
\object{HD~142527~A} is a Herbig Fe star surrounded by an inner disk, a gap of 120--130~au in radius seen both in sub-millimeter and scattered near-infrared (NIR) light, and an outer disk extending up to 700 au \cite[e.g.,][]{Fujiwara2006, Fukagawa2013, Avenhaus2017, Christiaens2014}.
The inner disk was shown to be inclined by $\sim 70\pm 5\degr$ with respect to the outer disk \citep{Marino2015}, and further evidence of this warp was found in the velocity map of CO lines observed with ALMA \citep{Casassus2015a}.
In addition to the gap, the outer disk reveals a horseshoe-shaped sub-millimeter continuum \citep{Ohashi2008, Casassus2013} and spiral arms seen in both $\mu$m-size dust and CO gas 
\citep[e.g.,][]{Fukagawa2006, Casassus2012, Rameau2012, Avenhaus2014, Christiaens2014}.
The age and mass of the star were estimated to $5.0\pm1.5$~Myr and $\sim$2~M$_{\sun}$ respectively, and the mass accretion rate was constrained to $2 \pm 1 \times 10^{-7} ~M_{\sun}$~yr$^{-1}$ \citep{Mendigutia2014,Lacour2016}. 
The parallax measured with Gaia corresponds to a distance of 156$\pm$6 pc \citep{Gaia2016}, which is consistent with the previous estimate of 140 $\pm$ 20 pc, based on proximity in the sky and similar proper motion with members of the Upper Centaurus Lupus moving group \citep{DeZeeuw1999, Teixeira2000, Fukagawa2006}. 
We adopt the Gaia distance throughout this work.

A low-mass companion around \object{HD~142527} was detected using sparse aperture masking \citep[SAM;][]{Biller2012}, and later confirmed with direct imaging
in the H$_{\alpha}$ line \citep{Close2014} and in $Y$ band \citep{Rodigas2014}.
HD~142527~B was found at $\sim 88$~mas ($\sim 14$~au) from the primary, at the inner edge of the large gap, hence probably 
shaping the warped inner disk.
A recent study suggested that the companion SED could be reproduced by a 3000-K companion with
an additional 1700-K circum-secondary disk component \citep{Lacour2016}.
Nevertheless, the study did not consider models with other values of effective temperature for the companion and its environment that might also reproduce the observed SED. 

More robust information on the nature of low-mass companions can be obtained through low- or medium-resolution spectra. 
Spectral-type classification of young M- and L-type objects can be performed either from comparison to spectral libraries or through the calculation of (gravity-independent) spectral indices \citep[e.g.,][]{Allers2013,Bonnefoy2014}.
Gravity (and hence age) is known to have a significant impact on the spectra of late-type objects \citep[e.g.,][]{Lucas2001,Allers2007,Cruz2009,Allers2013}. 
For young objects, spectral types can be converted into effective temperatures using a SpT-$T_{\mathrm{eff}}$ relationship that is intermediate between those of red giants and red dwarfs \citep[e.g.,][]{Luhman2003}.
Alternatively, the effective temperature of low-mass companions can also be estimated comparing the observed spectrum to grids of synthetic spectra generated with different atmospheric/photospheric properties \citep[e.g.,][]{Fortney2008, Allard2012}.
Using a Hertzsprung-Russel (HR) diagram to compare the effective temperature and absolute magnitude (or total luminosity) with stellar evolution tracks predicted from models allows us to estimate the mass and age of young stars \citep[e.g.,][]{Siess2000, Bressan2012, Baraffe2015}.
Finally, NIR spectra can also provide constraints on the mass accretion rate based on the observed intensity of hydrogen recombination lines \citep[e.g.,][]{Muzerolle1998, Calvet2004, Mendigutia2014}.


Here, we aim to better characterize 
low-mass companion \object{HD~142527~B}
using the integral field spectrograph VLT/SINFONI equipped with its pupil-tracking mode.
This analysis will enable to better assess the impact of the companion on the peculiar morphology of the disk in new hydro-dynamical simulations \citep{Price2018}.
In Sect.~\ref{obs+reduction}, we describe our observations and subsequent data reduction.
Section~\ref{specA} presents a brief analysis of the spectrum of the primary \object{HD~142527} A.
We then detail in Sect.~\ref{specB} the procedure to both extract the spectrum of \object{HD~142527} B and estimate photometric and astrometric uncertainties. 
The companion is analyzed in depth in Sect.~\ref{SpectrumAnalysis}, including fits of the spectrum to both synthetic and template spectra, spectral feature identification, and the use of evolutionary models to estimate physical properties of the companion. 
Results are discussed and compared to previous works in Sect.~\ref{discu}. 
Finally, Sect.~\ref{ccl} summarizes the main conclusions of this work.

\section{VLT/SINFONI observations and data reduction}\label{obs+reduction}

\begin{table*}[t]
\begin{center}
\caption{Summary of the observation of HD~142527 with VLT/SINFONI.}
\label{tab:Observations}
\begin{tabular}{lccccccccc}
\hline
Date & Filter & Pixel & UT time & NEXP$^{\mathrm{b}}$ & DIT$^{\mathrm{c}}$ & NDIT$^{\mathrm{d}}$ & Airmass & Seeing & Parallactic \\
 &  & scale$^{\mathrm{a}}$ & &  &  & &  &  & angle \\
& & (mas px$^{-1}$) & (Start/End) & & (s) & & (Start/End) & & (Start/End) \\
\hline
\hline
2014 May 10 & H+K & 12.5 & 04:35--06:40 & 40 & 1.5 & 30 & 1.07--1.07 & 0\farcs65--0\farcs70 & -34$\degr$/+38$\degr$\\
 & &  & 06:40--07:40 & 20 & 1.5 & 30 & 1.07--1.20 & 0\farcs90--1\farcs10 & +38$\degr$/+68$\degr$\\
& &  & 07:40--08:40 & 20 & 4.0 & 10 & 1.20--1.40 & $>$1\farcs10 & +68$\degr$/+83$\degr$\\
\hline
\end{tabular}
\end{center}
$^{\mathrm{a}}$ The y-axis has a twice lower sampling than the x-axis (i.e., its pixel scale is 25 mas px$^{-1}$ instead of 12.5 mas px$^{-1}$).\\
$^{\mathrm{b}}$ Number of data cubes acquired.\\
$^{\mathrm{c}}$ Detector integration time for each individual spectral cube.\\
$^{\mathrm{d}}$ Number of co-added spectral cubes in each data cube.\\
\end{table*}

VLT/SINFONI is an integral field spectrograph fed by an adaptive optics (AO) module \citep{Eisenhauer2003, Bonnet2004}.
As part of program 093.C-0526 (PI: S. Casassus), \object{HD~142527} was observed during the night of May 10, 2014, 
with SINFONI. The recently implemented pupil-tracking mode 
was used in order to take
advantage of angular differential imaging \cite[ADI; ][]{Marois2006}. 
Observations were made with the 0\farcs8 x 0\farcs8~field of view and the $H$+$K$ grating, offering a moderate spectral resolving power of $\sim$1500. 

The observing strategy consisted in a four-point dithering pattern simultaneous to the pupil-tracking, in order to sample the $\sim$1\arcsec~radius gap around the star as much as possible.
In practice, for each quartet of consecutive integrations, the star was placed close to a different corner of the detector.
A total of 80 data cubes were acquired. However, only the first 40 cubes, taken in photometric conditions, are used in this work.
They cover a total parallactic angle variation of $\sim$72$\degr$, and each cube consists of 30 co-added frames (NDIT) of 1.5s exposure time (DIT) each.
The seeing varied slightly between 0\farcs7~and 0\farcs9~for the first 40 cubes, then worsened significantly during the second half of the observation.
Considering the 60 first data cubes does not improve the results of Sects.~\ref{results} and \ref{SpectrumAnalysis}, and therefore we base our full analysis on the first 40 cubes only.
Details of the observations are summarized in Table \ref{tab:Observations}.

Data were first processed using the ESO pipeline (EsoRex version 3.10.2). This basic calibration included
dark subtraction, flat fielding, bad pixel removal, wavelength calibration, detector linearity correction, and extraction of spectral cubes from the raw frames. 
Each of our 40 spectral cubes is composed of 1992 spectral frames, spanning from 1.45~$\mu$m ($H$ band) to 2.45~$\mu$m ($K$ band). 
Each spectral frame is made of 64x64 pixels, with a horizontal plate scale of 12.5~mas per pixel,
and a vertical plate scale of 25~mas per pixel (i.e., each pair of consecutive rows has the same values).
Frames were squashed vertically down to 32 rows (by removing the redundant rows), and subsequently oversampled by a factor of two vertically
to match the horizontal and vertical pixel scales.

We noted that the bad pixels were not all well corrected by the ESO pipeline, so we applied our own bad-pixel correction algorithm appropriate for AO-corrected observations.
This algorithm replaced outlier values in concentric annuli centered on the star by the median value in the annulus with an additional random term proportional to the local noise.
At this point, the location of the star is different in 
each cube due to (i) the dithering pattern; 
(ii) the  atmospheric differential refraction effect on the different spectral channels  \citep[e.g.,][]{Roe2002}; (iii) the parallactic rotation (as the star was not placed at the center of the field of view); and
(iv) instrumental jitter. Therefore, we registered all the frames to place the star on the central pixel. 
As the star did not saturate during the observation, the exact location of the centroid was determined in each frame by fitting the observed point spread function (PSF) to a two-dimensional (2D) Gaussian.
Doing so in each spectral channel, we also estimated the full width at half maximum (FWHM) of the stellar centroid as a function of wavelength.

After this basic reduction, we built ADI cubes for each spectral channel, containing 40 frames each (from the 40 cubes taken in the best conditions),
and applied several ADI-based post-processing algorithms on each of them. 
The post-processing codes were adapted from the open-source Vortex Imaging Post-processing package\footnote{\url{https://github.com/vortex-exoplanet/VIP}} \citep[VIP;][]{GomezGonzalez2017}. 
Namely, we used (i) classical ADI, where the temporal median is subtracted from each frame pixel by pixel (Marois et al. 2006), (ii) ADI using principal component analysis \citep[PCA-ADI;][]{Amara2012, Soummer2012}
implemented on full frames (hereafter \emph{PCA-full}), (iii) PCA-ADI in several concentric annuli (hereafter \emph{PCA-annuli}), and (iv) PCA-ADI on a single annulus (hereafter \emph{PCA-annulus}).
PCA-full is similar to \emph{PynPoint} \citep{Amara2012}; the principal components are determined by singular value decomposition 
of a library consisting of all the ADI frames.
With PCA-annuli, each frame is divided into annuli of two~FWHM in width, paving at best the square frames.
This time the PCA library is built differently for each annulus, with a frame selection based on a given parallactic angle threshold, instead of admitting all the observed ADI frames.
We set this parallactic angle threshold to a minimum linear displacement of  0.5~FWHM at the radial separation of the companion.
This threshold acts to keep the frames where the companion would not have rotated sufficiently out of the PCA library, as these frames would lead to significant self-subtraction of the companion.
Finally, PCA-annulus performs PCA with no parallactic angle threshold on a single  annulus of 2 FWHM in width.
The latter method is much faster than the first ones, but needs a first estimate of the companion location to optimally define the annulus.

\section{Results}\label{results}

\subsection{The SINFONI spectrum of HD~142527~A}\label{specA}

\begin{figure*}[t]
\begin{center}
\sidecaption
\includegraphics[width=0.7\textwidth]{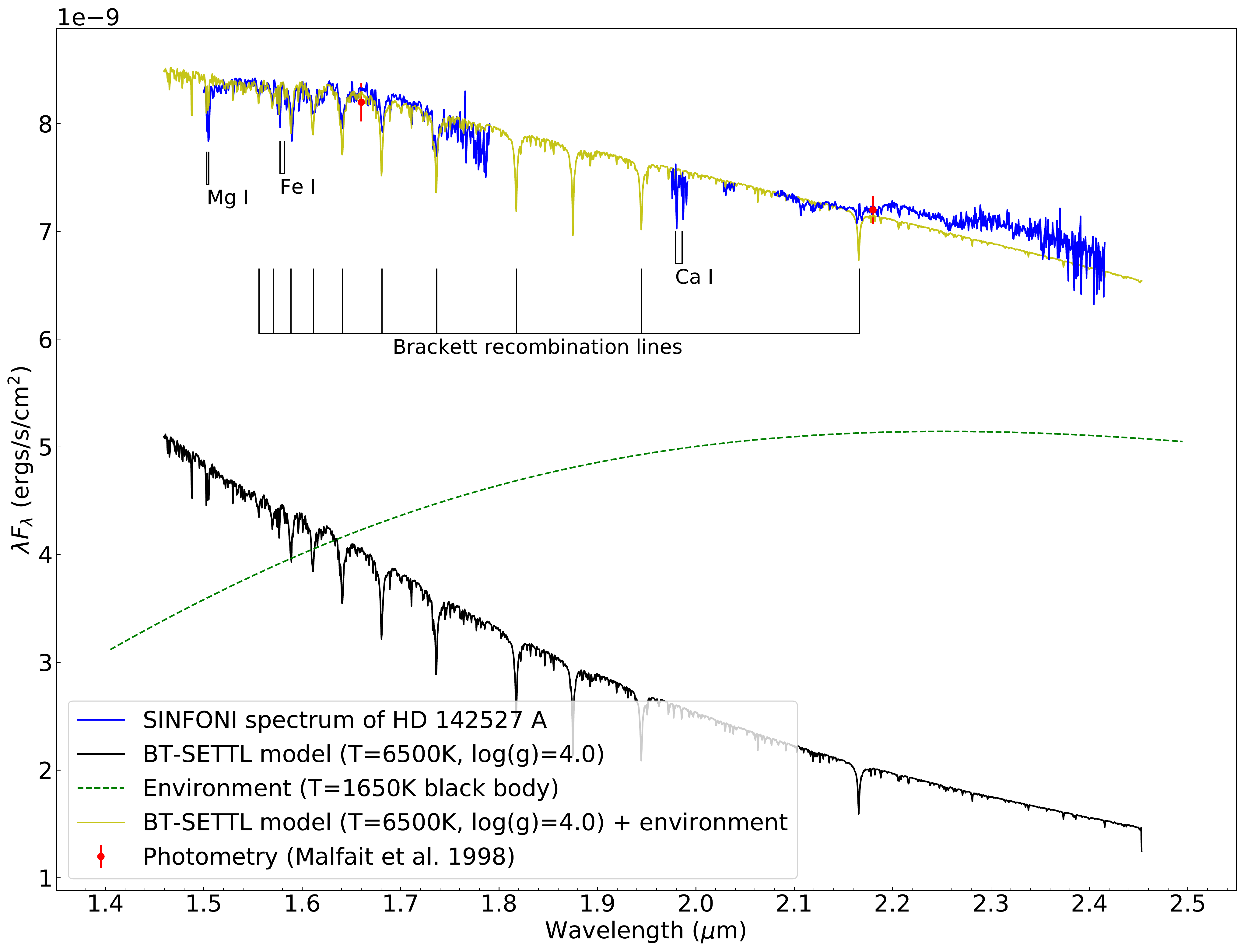}
\caption{\label{fig_spec_A}
Observed spectrum of HD~142527~A after telluric and instrumental response corrections (\emph{blue curve}). 
The \emph{yellow} curve corresponds to the combination of a BT-SETTL model ($T$=6500~K, log($g$)=4.0; \emph{black curve}) and a black body of 1650 K (\emph{green dashed curve}) representative of the hot inner disk rim.
Blanks in the SINFONI spectrum correspond to spectral channels where telluric lines were poorly corrected, and are therefore discarded for the rest of the analysis. 
}
\end{center}
\end{figure*}

We extracted the raw spectrum of  \object{HD~142527~A} with aperture photometry performed in each channel after basic calibration; that is, before ADI post-processing. 
The radius of the aperture was set to 0.5 FWHM, with the FWHM ranging from 4.9 to 6.4 pixels between the first and last spectral channel as a combined consequence of the power dilution of the PSF with wavelength and a poorer Strehl ratio at shorter wavelength.
This raw spectrum was then corrected for telluric lines.
Standard stars were observed before and after \object{HD~142527}, however either relative humidity or seeing were significantly different from their respective values during the observation of the science target.
Therefore, we used the \emph{molecfit} tool \citep{Smette2015, Kausch2015} for a refined correction of the telluric lines, based on fitting synthetic transmission spectra to our data.
 We also applied molecfit to the spectrum of the standard star observed in similar relative humidity but poorer seeing conditions to HD~142527. This allowed us to flag spectral channels where the telluric correction did not perform well and left significant residuals in the spectrum.
These channels were removed from the spectrum of \object{HD~142527~A} for the rest of the analysis.
The spectral slope of our observed spectrum was corrected from instrumental response using the model SED of \object{HD~142527~A} used in \citet{Casassus2015b}.

The final spectrum of \object{HD~142527~A}, provided in Fig.~\ref{fig_spec_A}, is consistent with the $H$- and $K$-band photometry reported in \citet{Malfait1998}.
We compare it to a model spectrum composed of a BT-SETTL model with $T$= 6500 K and log($g$)=4.0, and a dust/hot gas environment with a temperature of 1650 K with an emitting surface area of radius $0.10$--$0.15$ au. The environment temperature and physical extension are in agreement with constraints derived from modeling of NIR interferometry observations \citep{Lazareff2017}. The BT-SETTL model used for the primary is consistent with the best-fit temperature and gravity inferred from the VLT/XSHOOTER spectrum \citep{Mendigutia2014}. The BT-SETTL model was scaled using a radius of $3.20 R_{\odot}$, which is consistent with the expected radius of $3.19^{+0.41}_{-0.47} R_{\odot}$ based on the effective temperature of the star and its measured total luminosity \citep[$16.3 \pm 4.5 L_{\odot}$;][]{Mendigutia2014}. An extinction of $A_V$=0.8 was considered to redden the model \citep{Lazareff2017}. In order to be compared with our $H$+$K$ spectrum, our model is convolved with a Gaussian kernel with a size equal to the spectral PSF of SINFONI in the H+K mode, and smoothed to the spectral resolution of our SINFONI data (5~\AA~per channel).
The observed spectrum shows some excess emission at the red end of $K$-band with respect to the BT-SETTL + hot environment model, which can be explained by a combination of two factors: (i) the contribution of the inner (circumprimary) disk is more significant towards longer wavelengths, and (ii) the FWHM is larger in channels at the red end of the spectrum so that aperture photometry in those spectral channels includes more signal from the inner disk. 

Figure \ref{fig_spec_A} also labels the main spectral features identified in the VLT/SINFONI spectrum of \object{HD~142527~A}. 
The photospheric recombination lines of the Brackett series (Br14, Br13, Br12, Br11, Br10 and Br7 at 1.588, 1.611, 1.641, 1.681, 1.737 and 2.166 $\mu$m resp.) appear mostly veiled, based on the comparison between the observed spectrum and the BT-SETTL+hot environment model. 
This is an expected consequence from magnetospheric accretion \citep[e.g.,][]{Folha1999,Muzerolle2001}, which appears compatible with the significant mass accretion rate inferred in \citet{GarciaLopez2006} and \citet{Mendigutia2014}: $\sim 7 \times 10^{-8} ~M_{\sun}$~yr$^{-1}$ and $2 \pm 1 \times 10^{-7} ~M_{\sun}$~yr$^{-1}$, respectively.
The most significant \emph{emission line} is the strong Brackett-gamma
(Br$_{\gamma}$) line (2.16612~$\mu$m) that marginally stands out from
the underlying deep photospheric line. This line was already used by \citet{GarciaLopez2006} and \citet{Mendigutia2014} to estimate the mass accretion rate of \object{HD~142527~A}, based on observations of VLT/ISAAC and VLT/X-SHOOTER, respectively.
The shape of the observed line is in agreement with what would be obtained when degrading the X-SHOOTER spectrum to the spectral resolution of SINFONI.
In addition to the Brackett recombination lines, we also note the presence of absorption lines for the Mg~I triplet (1.503, 1.504 and 1.505 $\mu$m), Fe~I doublet (1.577 and 1.582 $\mu$m) and Ca~I doublet (1.979 and 1.986 $\mu$m) in the observed spectrum, which are all stronger than expected from the model.
This could suggest that we are witnessing additional absorption by refractory material in the close environment of the primary.

In Appendix \ref{app:HD142527A}, we describe our re-analysis of the VLT/XSHOOTER data used in \citet{Mendigutia2014}, which leads to a refined estimate of F6$\pm$0.5III-V for the spectral type of HD~142527~A.
This is compatible with the effective temperature estimates suggested both in \citet{Mendigutia2014} and by our comparison of the SINFONI spectrum to a BT-SETTL+hot environment model.
The physical characteristics of the primary, inferred both in previous works and from our SINFONI and XSHOOTER data analysis, are summarized in Table~\ref{tab:HD142527A}.

Comparison of our SINFONI spectrum to both the X-SHOOTER spectrum and a BT-SETTL+hot environment model consistent with previous literature indicates that the telluric correction and spectral calibration are valid. 
Nevertheless, for such low-resolution spectrum, the telluric correction is not reliable in some wavelength ranges dominated by atmospheric absorption which we choose to not show in Fig.~\ref{fig_spec_A} and discard for the rest of our analysis. 
We are left with 1313 
channels out of the 1992 original channels.

\begin{table}[t]  
\begin{center}
\caption{Characteristics of \object{HD~142527} and its close environment.}
\label{tab:HD142527A}
\begin{tabular}{lcc}
\hline
Parameter & Value & Ref. \\
\hline
\hline
\multicolumn{3}{c}{HD~142527~A}\\
\hline
Right ascension & $15^{\mathrm{h}}56^{\mathrm{m}}41^{\mathrm{s}}.89$ & \\
Declination & $-42\degr19\arcmin23\farcs5$  & \\
Spectral type & F6$\pm$0.5III--Ve & 1, 2 \\
$T_{\mathrm{eff}}$ [K] & $6500 \pm 100$ / $6550 \pm 100$ & 1, 3 \\
Log($g$) & $3.75 \pm 0.10$ & 3 \\
Luminosity [$L_{\odot}$] & $16.3 \pm 4.5$ & 3\\
Age [Myr] & $5.0 \pm 1.5$ & 3 \\
Mass [$M_{\sun}$] & $2.0 \pm 0.3$ & 3\\
Radius [$R_{\sun}$] & $3.2 \pm 0.2$ & 1\\ 
Distance [pc] & $156 \pm 6$ & 4 \\
$A_V$ [mag] & $0.60 \pm 0.05$ / $0.80 \pm 0.06$ & 5, 6\\
\hline
\multicolumn{3}{c}{Dust/hot gas environment}\\
\hline
Temperature [K] & $1650 \pm 50$ / $1680 \pm 100$ & 1, 6 \\
Radius [au] & 0.10--0.15 / $< 0.15$ & 1, 6\\ 
\hline
\end{tabular}
\end{center}
References: (1) This work; (2) \citet{Houk1978}; (3) \citet{Mendigutia2014}; (4) \citet{Gaia2016}; (5) \citet{Verhoeff2011}; (6) \citet{Lazareff2017}.
\end{table}

\subsection{Extraction of the spectrum of HD~142527~B}\label{specB}
\subsubsection{Re-detection of the companion}\label{redetection}

\begin{figure}[t]
\begin{center}
\includegraphics[width=0.49\textwidth]{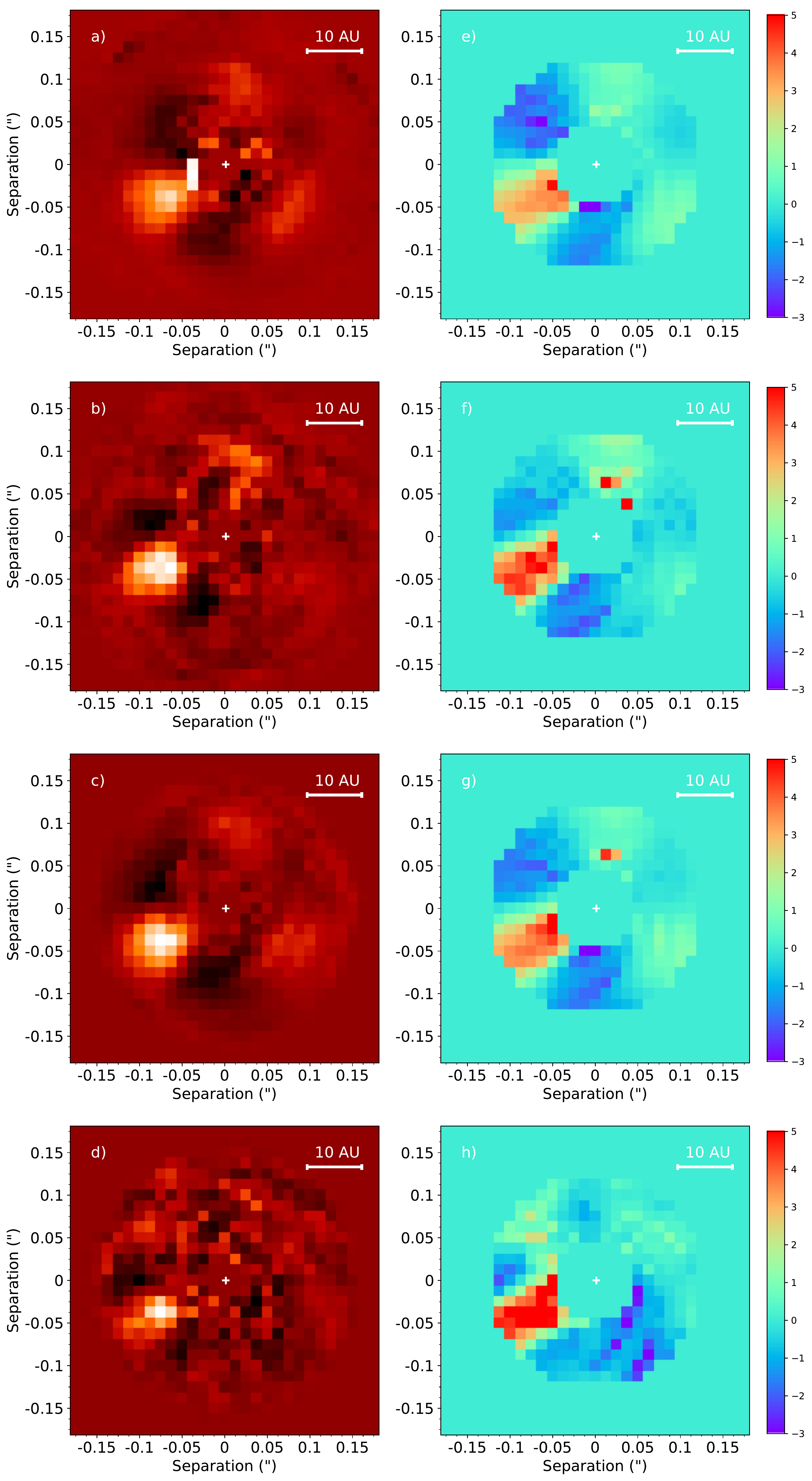}
\caption{\label{fig_conspicuous} Post-processing of the Br$_{\gamma}$ line spectral channel with {\bf a)} classical ADI;
{\bf b)} PCA-full with $n_{pc} = 7$; {\bf c)} PCA-annuli with $n_{pc} = 15$; {\bf d)} PCA-annulus with $n_{pc} = 10$.
{\bf e) - h)}  S/N map of a) - d) respectively. 
}
\end{center}
\end{figure}

The companion is re-detected in most spectral channels after the use of different ADI-based post-processing
algorithms. For illustration, the companion detection is shown for the spectral channel corresponding to the 
Br$_{\gamma}$ line in Fig.~\ref{fig_conspicuous}a, b, c, and d, respectively, for classical ADI,
PCA-full, PCA-annuli, and PCA-annulus (see Sect.~\ref{obs+reduction} for a description of the algorithms). The corresponding
signal-to-noise ratio (S/N) maps are provided in Fig. \ref{fig_conspicuous}e, f, g and h respectively.
The S/N was computed as in \citet{Mawet2014}, and therefore includes the small-sample statistics penalty.

\begin{figure}[t]
\begin{center}
\includegraphics[width=0.49\textwidth]{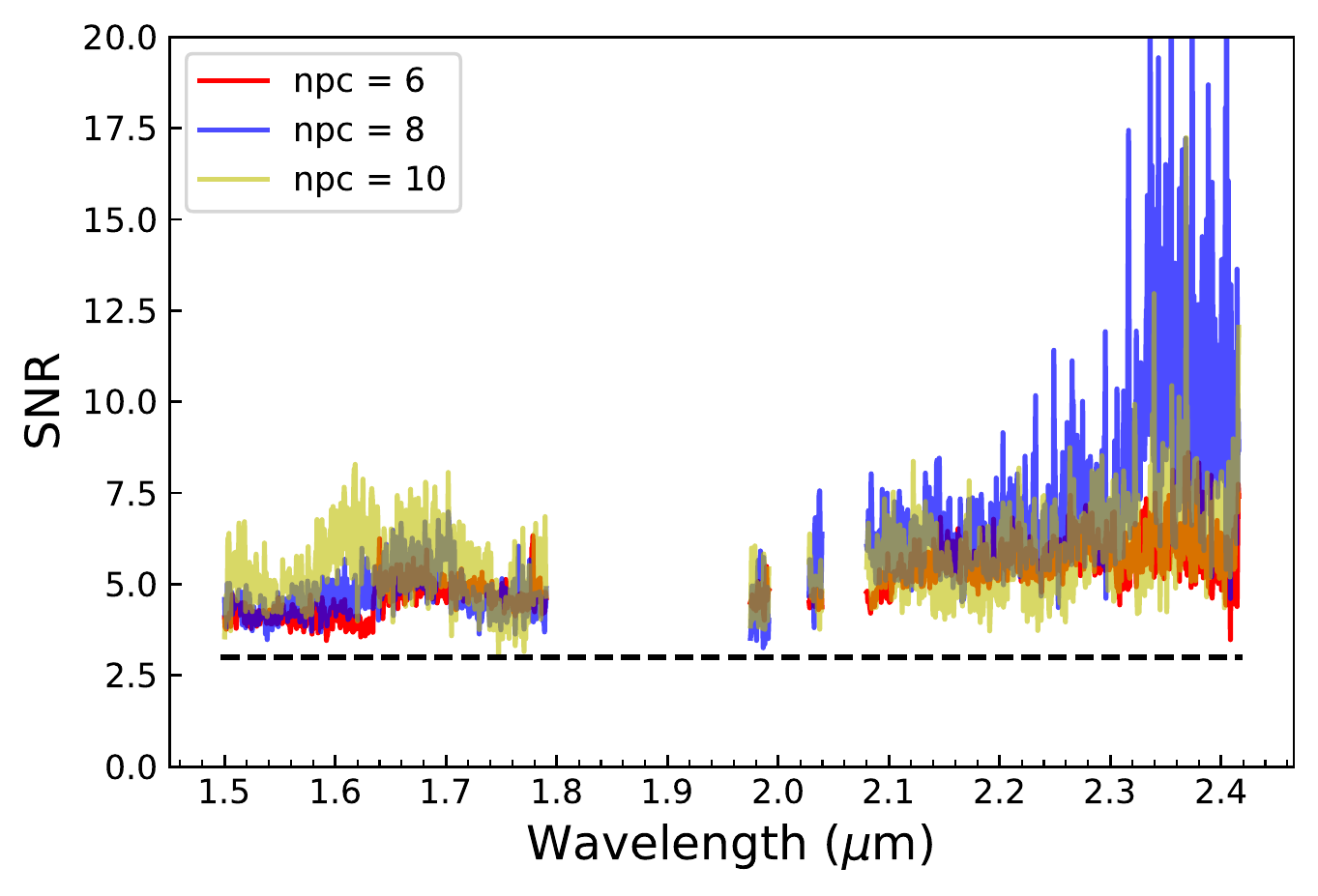}
\caption{ \label{fig_SNR}
S/N of the detection in each channel using PCA-annulus with 6, 8 and 10 principal components. The \emph{black dashed line} corresponds to a S/N of 3.
}
\end{center}
\end{figure}

For the spectral channel specific to the Br$_{\gamma}$ line, all post-processing methods provide a S/N $>$ 3 detection of the companion.
The PCA algorithms yield a higher S/N than classical ADI, although the exact S/N value depends on the 
number of principal components $n_{pc}$ used.
The companion is recovered with an S/N $\gtrsim$ 3 for $n_{pc}$ ranging from 1 to $\sim$20 
with either PCA-full or PCA-annuli, and for $n_{pc}$ ranging from 1 to 15 with PCA-annulus.
Using either classical ADI or PCA-annuli results in a higher residual speckle noise at the separation of the companion ($\sim$80~mas).
{}The speckle pattern in the frames created with classical ADI or PCA-annuli  for PSF subtraction is indeed intrinsically less correlated to the speckle pattern in the science frames than frames created with PCA-full or PCA-annulus.
As can be seen in Fig.~\ref{fig_conspicuous}a and c, the residual speckle pattern could be triangular with a possible contribution under the location of the companion.
For the rest of the analysis in this paper, we favor the PCA-annulus algorithm for ADI processing owing to its reduced computation time, and consider $n_{pc} \in [5,10]$ to reach a low residual speckle noise level while not oversubtracting the flux of the companion.
This choice is confirmed by our S/N estimates of the companion in all spectral channels. 
The detection is above 3$\sigma$ in all $H$+$K$ channels, and above 5$\sigma$ in most $K$-band channels (Fig.~\ref{fig_SNR}).


\subsubsection{Photometric and astrometric retrieval} \label{negfc}

\begin{figure}[th]
\begin{center}
\includegraphics[width=0.49\textwidth]{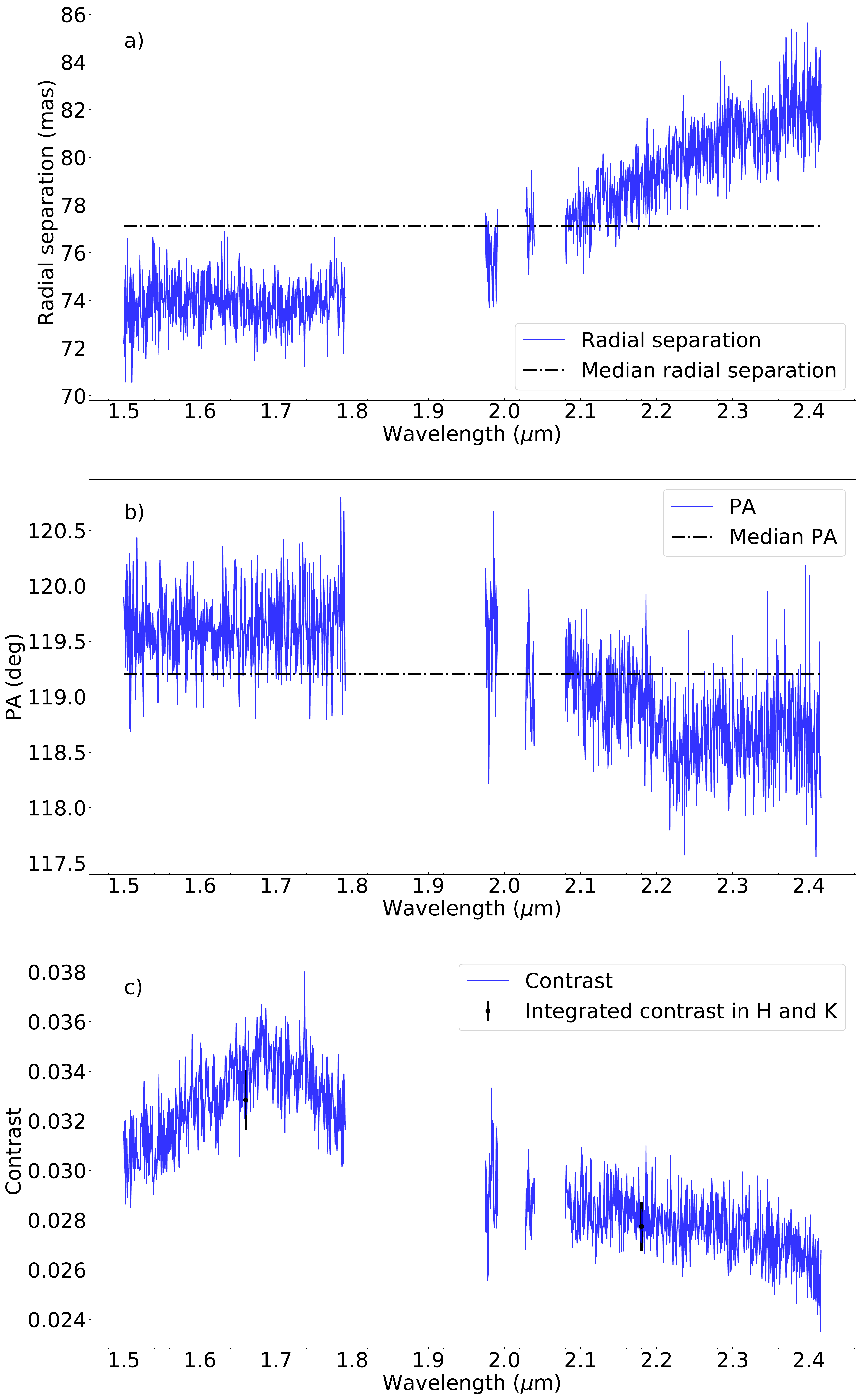} \\
\caption{ \label{fig_npc-SNR}
{\bf a)} Radial separation, {\bf b)} PA and {\bf c)} contrast of HD~142527~B in all $H$+$K$ channels, as determined with NEGFC.
The median separation and PA over all channels are indicated with \emph{dashed-dot lines}.
The contrasts integrated using the $H$- and $K$-band filter transmission curves of VLT/NACO, are provided with \emph{black error bars}.
}
\end{center}
\end{figure}

Motivated by the conspicuous detection of \object{HD~142527~B} in most spectral channels, 
we used the negative fake companion technique \citep[NEGFC;][]{Marois2010b,Lagrange2010} implemented in the VIP package in order to obtain an unbiased estimate
of the photometry and astrometry of the companion in all channels \citep[see details in ][]{Wertz2017}.
It is known that ADI algorithms affect the location and flux of any companion in the processed images.
The NEGFC technique circumvents this limitation with the injection of a negative fake companion in the frames
before they are processed with ADI.
For each spectral channel, we used the normalized median of the stellar PSF across the original 40 cubes as a template PSF for the fake companion to be injected with a negative flux.
At each wavelength, the method finds the optimal combination of radial separation $r_{\lambda}$, position angle PA$_{\lambda}$ and negative flux $-f_{\lambda}$ of \object{HD~142527~B} that
minimizes a specific figure of merit in the ADI-processed frame. 
The figure of merit consists usually in minimizing either the sum of absolute pixel intensities or the standard deviation of pixel values in a circular aperture centered on the companion location in the post-processed frame. 

As described in \citet{Wertz2017}, the optimization process of the figure of merit is performed in two steps (see the first two steps of their Sect.~3.2).
We first run a grid search on the companion flux, assuming a fixed companion position given by its highest pixel value.
The optimal flux and rough companion position are then used as input to run a more precise downward simplex algorithm 
using the same figure of merit but with three free parameters: $r_{\lambda}$, PA$_{\lambda}$ and $-f_{\lambda}$.

The choice of parameters associated to NEGFC, such as the number of principal components to be used, the size of the aperture and the figure of merit to be optimized, is crucial to obtain reliable results.
We noticed that using a single set of parameters led to a significant amount of outliers in estimated $r_{\lambda}$, PA$_{\lambda}$, and $f_{\lambda}$ throughout the 1313 spectral channels of our spectrum. 
Therefore, we considered different sets of NEGFC parameters within reasonable ranges, constrained as  follows.
\begin{itemize}
\item For the post-processing ADI algorithm, we chose PCA-annulus, owing to its time-efficiency, with $n_{pc} \in [5,10]$ as it optimizes the S/N of the companion (Fig.~\ref{fig_SNR}). This range of $n_{pc}$ also provides a visually low residual speckle noise level while preserving the flux of the companion from significant oversubtraction.
\item We considered two different figures of merit: minimizing either the sum of absolute pixel intensities or the standard deviation of pixel values in a circular aperture centered on the companion location in the post-processed frame. 
\item We used 0.7 and 0.9 FWHM-radius circular apertures for the minimization of the figure of merit. These choices are justified by (i) the requirement for the aperture radius to always be shorter than the radial separation of the companion ($\sim$ 6.4 pixels, or $\sim$ 1 FWHM at the longest wavelength) and (ii) the need for a sufficient number of pixels in the aperture for the figure of merit to be statistically meaningful and therefore enable convergence to the correct solution. 
\end{itemize}

For each spectral channel, we considered the median parameters of the companion derived using the $6\times2\times2 = 24$ different sets of NEGFC parameters.
The median enabled us to discard all outlier results, which suggests that NEGFC converges in general to a consistent solution in terms of $r_{\lambda}$, PA$_{\lambda}$, and $-f_{\lambda}$, but also that particular combinations of aperture size, figure of merit, and number of principal components can sometimes lead to significant contamination by residual speckle noise.
The results are reported in Fig.~\ref{fig_npc-SNR}.
The radial separation and PA of the companion derived in each spectral channel are compared to their median value over all spectral channels.
The flux of \object{HD~142527~B} is expressed as a contrast with respect to the primary star.

In Fig.~\ref{fig_npc-SNR}a, we notice that $r_{\lambda}$ seems to increase slightly with wavelength in $K$ band (by up to $\sim$10~mas), which is reminiscent of the behavior of a speckle.
However, a speckle at that radial distance would move radially by $\sim$50~mas from 1.45~$\mu$m to 2.45~$\mu$m, and would also be expected to show a more monotonic trend.
Nevertheless, a putative speckle located at 60~mas at the short end of $H$ band would lie at less than 1~FWHM from the companion, and would still be unresolved from the companion at the long end of the $K$ band, possibly shifting the centroid of the companion depending on its flux ratio.
Another possibility is that the companion emission is radially extended (outward) with a stronger contribution at longer wavelengths.
This feature was already noted by \citet{Rodigas2014} with polarimetric observations, which could indicate the presence of heated material at larger separation from the companion, shifting the centroid accordingly.
The PA variation with wavelength (Fig.~\ref{fig_npc-SNR}b) appears intimately related to $r_{\lambda}$, in particular at longer wavelengths, where PA$_{\lambda}$ appears to decrease slightly while $r_{\lambda}$ increases (Fig. \ref{fig_npc-SNR}b).

The median values of radial separation and PA of \object{HD~142527~B} over all $H$- and $K$-band channels 
are 77.1$\pm$3.3~mas and 119.2$\pm$0.6$\degr$, respectively.
The quoted uncertainties are the standard deviation of $r_{\lambda}$ and PA$_{\lambda}$ over all channels, respectively. 
More robust estimates of the uncertainty on these values are provided in Sect. \ref{errors}.

Regarding the contrast ratio with respect to the primary (Fig.~\ref{fig_npc-SNR}c), we notice a triangular shape throughout the $H$ band with values ranging between 0.030 and 0.036. 
The contrast ratio in $K$ band shows a monotonically decreasing trend from 0.030 to 0.025 with increasing wavelength.
We integrated the derived contrasts using the $H$- and $K$-band filter transmission curves of VLT/NACO, and found that the $H$-band (resp. $K$-band) contrast is $3.28 \pm 0.12 \times 10^{-2}$ (resp. $2.78 \pm 0.10 \times 10^{-2}$).

\subsubsection{Estimation of the uncertainties}\label{errors}


At such small angular separation from the primary ($\sim$ 77~mas), the residual speckle noise is expected to be the largest source of uncertainty on the radial separation, PA, and contrast of HD~142527~B derived by NEGFC.
The procedure that we followed to estimate the residual speckle noise uncertainties on the parameters of the companion in each spectral channel is described in Appendix~\ref{app:uncertainties}.
We used a weighted mean over all spectral channels to estimate the final radial separation and PA of the companion.  
This leads to $r = 6.24 \pm 0.14$ px ($r = 78.0 \pm 1.7$ mas) and PA $= 119.1 \pm 0.8 \degr$. 

Nervertheless, a complete astrometric error budget should not only consider the residual speckle noise uncertainties, but also (i) the error associated to the centering of the star in the frames, (ii) the error related to the plate scale ($\sim 12.5$~mas per pixel), and (iii) the errors related to true north and pupil offset \citep{Wertz2017}. 
The stellar PSF is not saturated in any of our data. 
Therefore, the centering of the star was simply performed with a Gaussian fit of the centroid. 
Comparison with a Moffat fit in all spectral channels provides an agreement of $\sim$0.05~pixel. 
We do not expect the error associated to stellar centering to be much larger. 
It has to be noted that we lack appropriate observations to derive a proper estimate of the errors associated to plate scale, true north, and pupil offset.
However, \citet{Meshkat2015} quoted total uncertainty values (including true north) of 0.4~mas and 0.5$\degr$~ for $r$ and PA for another low-mass companion detected by SINFONI in pupil tracking.
The latter companion was found further away from its central star, and therefore relatively free from speckle noise contamination.
As our companion lies at much closer separation, the term associated to the plate scale uncertainty (proportional to $r^2$) in the error budget is expected to be negligible.
Their observations were only a few months apart from ours, so we can assume similar errors on true north and pupil offset as theirs.
We conservatively consider their uncertainties and sum them in quadrature to our residual speckle uncertainties.
Our final astrometry is therefore the following: $r = 78.0 \pm 1.8$~mas and PA~$= 119.1 \pm 1.0\degr$.
These values are in agreement with the expected position given by MagAO and GPI data acquired at the same epoch \citep{Rodigas2014, Lacour2016}.

Our tests in Appendix~\ref{app:uncertainties} suggest that an uncertainty of up to 15\% could affect the estimated absolute contrast of the companion if it lies on top of a speckle feature similar to the second brightest artefact (after the companion itself) seen at the same radial separation (Fig.~\ref{fig_conspicuous}).
However, we also estimated a relative uncertainty $\lesssim 5$\% regarding the shape of the contrast spectrum, even when on top of such a speckle feature.
This suggests that a spectrum of good quality can be extracted for the companion.

\subsubsection{Spectrum of HD 142527 B} \label{final_spec_B}

\begin{figure*}[tbh]
\begin{center}
\includegraphics[width=\textwidth]{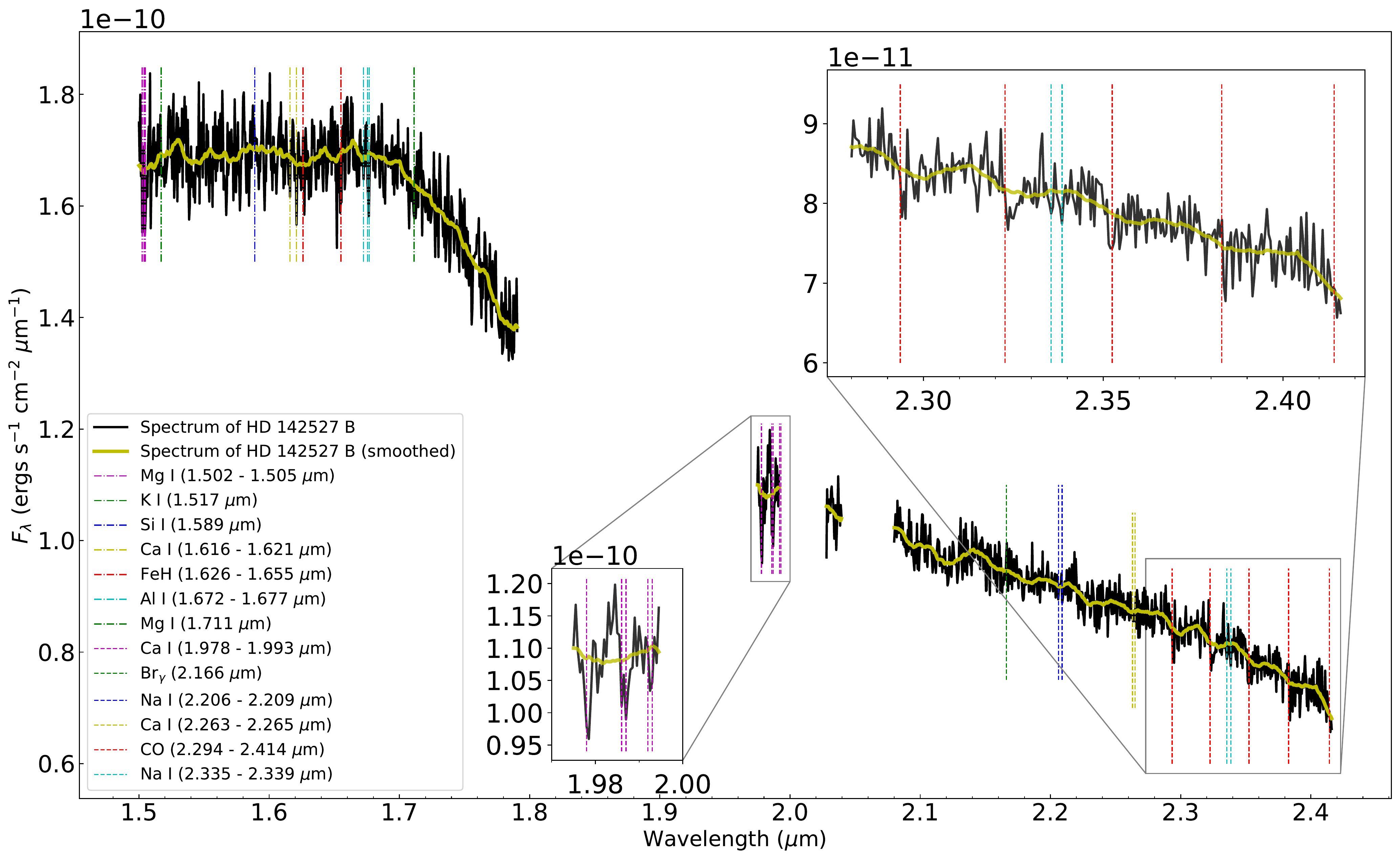}
\caption{ \label{fig_spec_B_lines}
Observed $H$+$K$ spectrum of HD~142527~B (\emph{black curve}) and spectrum after Savitzky-Golay filtering (\emph{yellow curve}), along with expected lines in the spectrum of an early to mid-M dwarf. Lines in $H$- and $K$-band are provided with \emph{dashed-dot} and \emph{dashed lines}, respectively. The combined presence of five tentative jumps at expected CO transitions (2.294, 2.323, 2.352, 2.383 and 2.414 $\mu$m) makes the detection of the first overtone of CO bandhead significant. The Ca I quintuplet (1.978, 1.986, 1.987, 1.992 and 1.993 $\mu$m) also appears to be detected, although this area is bordered by poorly corrected telluric features (not shown) which inspire caution. Insets zooming on the Ca I quintuplet and CO bandhead spectral regions are provided.} 

\end{center}
\end{figure*}

The spectrum of the companion, given in Fig. 5, is simply obtained by multiplying the spectrum of \object{HD~142527~A} (Fig.~\ref{fig_spec_A}) by the contrast ratio found in each spectral channel (Fig.~\ref{fig_npc-SNR}c).
Both the absolute and relative flux uncertainty on the spectrum of HD~142527~A are negligible compared to the uncertainty on the contrast of HD~142527~B (Sect.~\ref{errors}).
The absolute flux uncertainty of the primary is very small given the consistency between our spectrum and $H$- and $K$-band photometric measurements \citep[][Fig. \ref{fig_spec_A}]{Verhoeff2011}.
The relative flux uncertainty on the spectrum of HD~142527~A is related to the extraction of the spectrum using aperture photometry in each spectral channel.
Based on Poisson statistics, we estimate the latter to be $\lesssim$0.5\% in all channels.
Therefore, we only consider the uncertainties related to residual speckle noise in the rest of this work, since it is at least an order of magnitude larger than the uncertainties related to the spectrum of HD~142527~A. 

We applied a Savitzky-Golay (SG) filter \citep{Savitzky1964} with a window size of 51 channels and a polynomial of order 3 to smooth our spectrum. 
This is shown with the \emph{yellow curve} in Fig.~\ref{fig_spec_B_lines}. 
We compared the SG-filtered spectrum and the spectrum after a weighted binning of factor 6 (where the weight is inversely proportional to the contrast uncertainty), and noted that the resulting curves are consistent with each other. 
We favor the SG-smoothing, as it does not suffer from a loss in wavelength resolution and is known to be more robust to outliers than channel binning \citep{Savitzky1964}.


\section{Characterization of HD~142527~B}\label{SpectrumAnalysis}

The spectrum of HD~142527~B shows a relatively flat $H$-band continuum shortward from 1.7 $\mu$m followed by a steep drop, turning into a moderate negative slope in $K$-band.
This shape contrasts with the $H+K$ spectrum of young brown dwarfs (spectral type later than M6), characterized by a triangular-shaped $H$-band spectrum peaking at $\sim$1.67 $\mu$m and a hump in $K$ band centered on 2.25$\mu$m, which are the results of broad water absorption bands \citep{Jones1994,Lucas2001,Reid2001,Luhman2004,McGovern2004}.
The absence of these easily identifiable features suggests that HD~142527~B is of earlier spectral type than M6.
In Fig.~\ref{fig_spec_B_lines}, we show lines expected to be present in the $H$+$K$ spectrum of early to mid-M dwarfs. 
These will be discussed in Sect.~\ref{SpectralFeatures}.
Hereafter, we carry out an in-depth analysis of the spectrum of the companion in order to better constrain its spectral type, effective temperature, surface gravity, mass and age. 

\subsection{Fit with BT-SETTL models}\label{BTSETTL}

In order to interpret the $H$+$K$ spectrum of \object{HD~142527~B}, we first compare it to a set of BT-SETTL synthetic spectra\footnote{\url{http://perso.ens-lyon.fr/france.allard/}} \citep{Allard2012}. 
BT-SETTL models are available for a large range of temperatures and surface gravities.
These synthetic spectra are given in flux units at the stellar surface.
As a consequence, one must assume a certain distance, stellar radius, and extinction to be able to compare the models with our observations.
Assuming a distance of 156~pc \citep{Gaia2016}, only the stellar radius and extinction are left as free parameters.
We choose to not fix the extinction to any value estimated for the primary given the possibility that the companion is self-embedded or surrounded by an optically thick circum-secondary disk.
Models are considered on a grid of four free parameters: effective temperature, surface gravity, stellar radius and extinction.
In order to reflect the possible NIR contribution of hot circumsecondary material, we considered a second fit including two additional parameters: the radius and temperature of a hot inner rim \cite[see e.g.,][]{Cieza2005}.
Free parameters and related assumptions are detailed in Sect.~\ref{free_params}.

Although photometric measurements of the companion are available at other wavelengths, we chose to fit the BT-SETTL models to our H+K spectrum alone.
This is motivated by the fact that (i) an arbitrary choice of relative weight given to the spectroscopic and photometric points might change the best-fit result, and  that (ii) flux calibrations with different instruments used at separate wavelengths could introduce unpredictable biases in the results.
In order to be compared with our $H$+$K$ spectrum, the BT-SETTL models are convolved with a Gaussian kernel with a size equal to the spectral PSF of SINFONI for the mode we used, and smoothed to the spectral resolution of our observations (5~\AA~per channel).

\subsubsection{Free parameters}\label{free_params}

\paragraph{Effective temperature}
The parameter with the most impact on the shape of the spectrum is the effective temperature of the companion ($T_{\mathrm{eff}}$). Its value is intimately related to the slope of the model spectrum between 1.7 and 2.4 $\mu$m.
The BT-SETTL grid of synthetic spectra ranges from 1200~K to 7000~K in $T_{\mathrm{eff}}$, with steps of 100~K.
Based on visual similarity of spectral slopes between data and model and on the previous estimate of 3000~K from \citet{Lacour2016}, we restricted the grid used for the fit to temperatures between 2600 and 4500~K.

\paragraph{Surface gravity}
The surface gravity ($\log(g)$) has an impact on the \emph{cuspiness} of the spectrum at around $1.7~\mu$m.
This is related to water vapor absorption bands shortward of 1.55 $\mu$m and longward of 1.72 $\mu$m being more significant at lower gravity \citep[e.g.,][]{Allers2007}.
We tested values of surface gravity from 2.5 to 4.5 with steps of 0.5, thereby covering the range in values of $\log(g)$ corresponding to giants and main sequence stars.

\paragraph{Stellar radius}
The BT-SETTL synthetic spectra are provided in units of emitted flux at the stellar surface. Therefore, the stellar radius $R_B$ is adjusted to scale each model to units of observed flux through the dilution factor ($R_B^2/d^2$), where the distance $d$ is set to 156~pc.
Our grid included 150 values of $R_B$ geometrically spaced between 0.1 and 10.0~$R_{\sun}$. 

\paragraph{Extinction}
We assumed that extinction can be characterized by only one parameter: $A_H$, the extinction in $H$ band. 
This value is then extrapolated to other wavelengths using a total-to-selective extinction ratio R$_V$ = 3.1 following \citet{Draine1989}.
The extinction that is applied to the different models is adjusted to match the global slope of the continuum from $H$ to $K$ band. 
We tested values of $A_H$ between 0 and 1.2, in steps of 0.05, thereby including the values of $A_V =0.6$ ($A_H \sim0.1$), $A_V =0.8$ and $A_H = 0.3$ suggested in \citet{Verhoeff2011}, \citet{Lazareff2017} and \citet{Close2014}, respectively.

\paragraph{Circum-secondary disk}
In view of both the young age of the system and the detection of H$_{\alpha}$ emission at the location of \object{HD~142527~B} \citep{Close2014}, it is conceivable that the companion is surrounded by a circum-secondary disk.
Its inner rim could be at sufficiently high temperature to contribute significantly to the NIR flux \citep[e.g.,][]{Cieza2005}.
We assumed for simplification that the latter could be represented by a uniform disk emitting as a black body, therefore requiring only two parameters for its characterization: the disk radius $R_d$ and temperature $T_d$.
We tested values of temperature ranging from 1000~K to 1700~K in steps of 100~K, with 22 values of radii geometrically spaced between 0.005 and 0.100 au ($\sim$ 1.5 to 20 $R_{\sun}$). 
We are not sensitive to temperatures lower than $\sim$ 1000~K, since they do not lead to a significant contribution in $H$ or $K$ band.  
Temperatures above 1700~K are not considered as the dust would be likely sublimated \citep[see e.g.,][]{Meyer1997}.
We insist on the fact that this is a very simplistic representation of the inner rim, which aims to answer qualitatively the question of whether the observed spectrum suggests the presence of an additional component (apart from the photosphere) or not.
A more realistic model would require consideration of an annular geometry and would therefore involve three free parameters (the temperature, radial separation, and width of the annulus), where the last two parameters would not be independently constrained by our observations.

\subsubsection{Best fit BT-SETTL model} \label{bestfit_BTSETTL}

\begin{figure*}[htb]
\begin{center}
\includegraphics[width=\textwidth]{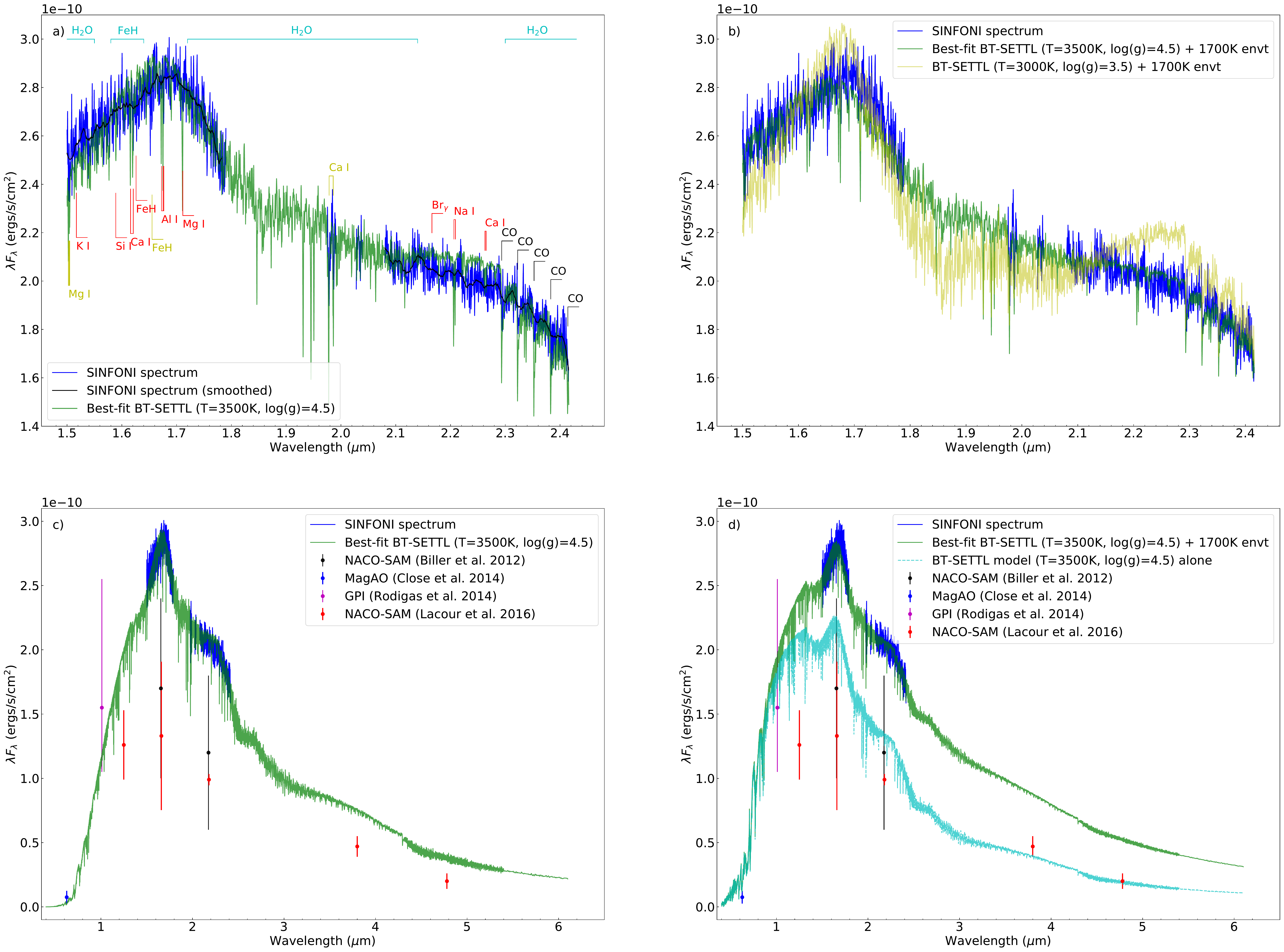}
\caption{ \label{fig_spec_B_modeling}
{\bf a)} Observed $H$+$K$ spectrum of HD~142527~B (\emph{blue curve}) and spectrum after Savitzky-Golay filtering (\emph{black curve}). The best-fit BT-SETTL synthetic spectrum is obtained with $T_{\mathrm{eff}} = 3500$ K and $\log(g)=4.5$ (\emph{green curve}). For reference, we provide the location of spectral lines that are expected in the H+K spectrum of M-type stars. Lines labeled in black, yellow, and red are detected, tentatively detected, and non-detected in the SINFONI spectrum, respectively.
{\bf b)} The observed $H$+$K$ spectrum of HD~142527~B (\emph{blue curve}) is compared to the best-fit BT-SETTL+black body model, representing the photospheric and hot circum-secondary environment contributions, respectively. The best fit is obtained with a $T_{\mathrm{eff}} = 3500$ K and $\log(g)=4.5$ BT-SETTL model, with a $1700$~K environment (\emph{green curve}). We also plot the model suggested in \citet{Lacour2016} consisting of a $T_{\mathrm{eff}} = 3000$ K and $\log(g)=3.5$ BT-SETTL model, with a $1700$ K environment (\emph{yellow curve}).
{\bf c)} As in a) but for a larger wavelength range. 
{\bf d)} As in b) but for a larger wavelength range, but without the \citet{Lacour2016} model and showing the contribution of the photosphere of the companion (\emph{cyan} curve) in our best-fit BT-SETTL+environment model. In both c) and d), we compare our spectrum with photometric measurements of the companion obtained with different instruments: VLT/NACO-SAM \citep[\emph{black points;}][]{Biller2012}, MagAO \citep[\emph{blue points;}][]{Close2014}, GPI \citep[\emph{magenta points;}][]{Rodigas2014} and VLT/NACO-SAM \citep[\emph{red points;}][]{Lacour2016}. 
}
\end{center}
\end{figure*}

We assume that using the $\chi^2$ metric as a goodness-of-fit estimator can provide a reasonable estimate of the effective temperature, surface gravity, extinction and stellar radius. 
Nevertheless, this estimator is somewhat flawed as the points of our spectrum are not statistically independent  -- they are significantly affected by the same speckle noise scaled by $\lambda$ -- and do not rigorously follow a Gaussian error distribution \citep{Soummer2007,Greco2016}. 
Bootstrapping \citep[e.g.,][]{Hastie2009} was therefore implemented to estimate the uncertainties on each of the best-fit parameters due to spectral correlation of errors between different channels.
We created 5000 bootstrap samples, by drawing 1313 random channels, with repetition, following a uniform probability distribution among our 1313 channels. 
The first of these samples consists of the 1313 good channels of the actual observed spectrum.
For each bootstrap sample, we find the model of our grid that minimizes the residuals, and save the corresponding model parameter values.
Histograms reporting the occurrence of each parameter value are then fitted to Gaussian functions in order to estimate the 1$\sigma$--uncertainty on each parameter. 

For $R_B$ and $R_d$, we propagate an additional uncertainty reflecting the $\sim$15\% uncertainty on the absolute flux of the companion (see Appendix~\ref{app:uncertainties} and Fig.~\ref{fig_speck_errors} for details).
This is added in quadrature to the uncertainty on these parameters estimated with bootstrapping.
Since $T_{\mathrm{eff}}$, $\log(g)$, $A_H$ and $T_d$ control the shape of the spectrum, that is, the channel to channel relative flux, but not the absolute flux, these parameters are not affected by the additional uncertainty on the absolute flux.

The best-fit parameters for the companion and a putative hot inner rim, 
are compiled in Table \ref{tab:best_fit_params}. 
We distinguish between models consisting of photospheric emission alone (4 free parameters), and of the sum of photospheric and hot circum-secondary material signals (2 additional free parameters), in the top and bottom parts of Table \ref{tab:best_fit_params}, respectively.


\begin{table}  
\begin{center}
\caption{Best fit parameters for HD~142527~B and a putative hot inner rim.}
\label{tab:best_fit_params}
\begin{tabular}{lcc}
\hline
Parameter & Searching range & Best fit value \\
\hline
\multicolumn{3}{c}{Companion alone}\\
\hline
$T_{\mathrm{eff}}$ [K] & 2600--4500 & $3500 \pm 100$ \\
$\log(g)^{\dagger}$ & 2.5--4.5 & $4.5_{-0.5}$ \\ 
$R_B$ [$R_{\sun}$] & 0.1--10.0 & $2.08 \pm 0.18 $ \\ 
$A_H$ [mag] & 0.0--1.2 & $0.75^{+0.05}_{-0.10}$\\
$\chi_r^2$ & & 0.14 \\ 
\hline
\multicolumn{3}{c}{Companion + hot inner rim}\\ 
\hline
$T_{\mathrm{eff}}$ [K] & 2600--4500 & $3500 \pm 100$ \\
$\log(g)^{\dagger}$ & 2.5--4.5 & $4.5_{-0.5}$ \\ 
$R_B$ [$R_{\sun}$] $^{\ddagger}$& 0.1--10.0 & 1.09--1.55 (1.42) \\ 
$A_H$ [mag] $^{\ddagger}$ & 0.0--1.2 & 0.0--0.2 (0.2) \\ 
$T_d$ [K] $^{\ddagger}$& 1000--1700 & 1500--1700 (1700) \\ 
$R_d$ [$R_{\sun}$] $^{\ddagger}$& 1.5--20.0 & 3.0--13.9 (11.8) \\ 
$\chi_r^2$ & & 0.10 \\ 
\hline
\end{tabular}
\end{center}
$^{\dagger}$Missing upper uncertainty limit reflects the fact that the best-fit parameter is found at the limit of the searching range.\\
$^{\ddagger}$Given the parameter degeneracy between $R_B$, $A_H$, $T_d$ and $R_d$, we provide parameter ranges for which a similar-quality fit can be obtained ($\chi_r^2 \sim 0.10$). Values in parentheses correspond to the model shown in Fig.~\ref{fig_spec_B_modeling}b and d.



\end{table}

Fitting the whole spectrum with pure photospheric signal (captured by the BT-SETTL synthetic spectra) yields a best-fit effective temperature and surface gravity of $T_{\mathrm{eff}} = 3500 \pm 100$~K and $\log(g) = 4.5_{-0.5}$, respectively ($\chi_r^2 = 0.14$).
The corresponding best-fit companion radius $R_B$ and extinction $A_H$ are $R_B = 2.08 \pm 0.18 R_{\sun}$ and $A_H = 0.75^{+0.05}_{-0.10}$, respectively.
The best-fit BT-SETTL synthetic spectrum is shown in Fig.~\ref{fig_spec_B_modeling}a and c, in units of $\lambda F_{\lambda}$ to better show small changes in spectral slopes; it accounts relatively well for the $H$-band spectrum of the companion. 
The $K$-band spectrum is also qualitatively comparable, although the slope is not perfectly reproduced; either too flat at 2.15--2.28 $\mu$m or too steep beyond 2.33 $\mu$m.


Adding two additional free parameters representing hot circum-secondary material appears to slightly improve the quality of the fit (best-fit $\chi_r^2 = 0.10$).
The best-fit effective temperature and surface gravity are consistently $T_{\mathrm{eff}} = 3500 \pm 100$~K and $\log(g) = 4.5_{-0.5}$.
However, the two additional free parameters introduce a degeneracy between the other parameters, as different combinations of $R_B$, $A_H$, $T_d$ and $R_d$ can produce very similar quality fits. 
Therefore, we prefer to provide best-fit parameters $R_B$, $A_H$, $T_d$ and $R_d$ as ranges of values, reflecting the different combinations of these parameters leading to $\chi_r^2 \approx 0.10$:
$R_B \in [1.09, 1.55] R_{\sun}$, $A_H \in [0.0, 0.2]$,  $T_d \in [1500, 1700]$~K and $R_d \in [3.0, 13.9] R_{\sun}$.
Among these models, we slightly favor the one associated to $A_H = 0.2$, as it leads to a closer $R$-band flux to the photometric measurement in \citet{Close2014}.
This model is shown with the \emph{green curve} in Fig.~\ref{fig_spec_B_modeling}b and d and corresponds to the set of best-fit parameters given in parentheses in Table~\ref{tab:best_fit_params}.
Nevertheless, we also note that the best-fit BT-SETTL model without inclusion of a hot environment still leads to a slightly better agreement in $R$-band (Fig.~\ref{fig_spec_B_modeling}c). 
This is due to the higher value for the best-fit extinction ($A_H = 0.75^{+0.05}_{-0.10}$).

The H+K spectrum is globally better reproduced when allowing for the additional contribution of a hot black-body environment. 
For comparison, we also plot the 3000~K BT-SETTL + 1700~K circum-secondary environment model suggested in \citet{Lacour2016} to fit the SED of the companion (\emph{yellow curve} in Fig.~\ref{fig_spec_B_modeling}b).
It appears incompatible with our SINFONI spectrum, given its more pointy shape $H$-band continuum and the hump centered at $\sim 2.25~\mu$m that are both typical of later-type objects.
We also show the contribution of the photosphere alone in our best-fit BT-SETTL + hot environment model (\emph{cyan curve} in Fig.~\ref{fig_spec_B_modeling}d). 
This will be relevant for the estimation of $H$- and $K$-band magnitudes of the companion (without the contribution of the circum-secondary disk), that will be used to locate the position of the companion in HR diagrams.

Figure \ref{fig_spec_B_modeling}c and d compares our best-fit models to the whole SED of the companion, obtained from previous detections with different instruments.
Both the best-fit BT-SETTL model alone and best-fit BT-SETTL+environment model are roughly consistent with the $R$-band and $Y$-band measurements with MagAO and GPI, respectively \citep{Close2014,Rodigas2014}.
However, our observed spectrum and best-fit models are significantly brighter than photometric measurements of the companion obtained using SAM with VLT/NACO \citep{Biller2012,Lacour2016}. 
This is further discussed in Sect.~\ref{SEDfitting}.


\subsection{Fit with template spectra}\label{realspectra}

\begin{figure}[tb]
\begin{center}
\sidecaption
\includegraphics[width=0.49\textwidth]{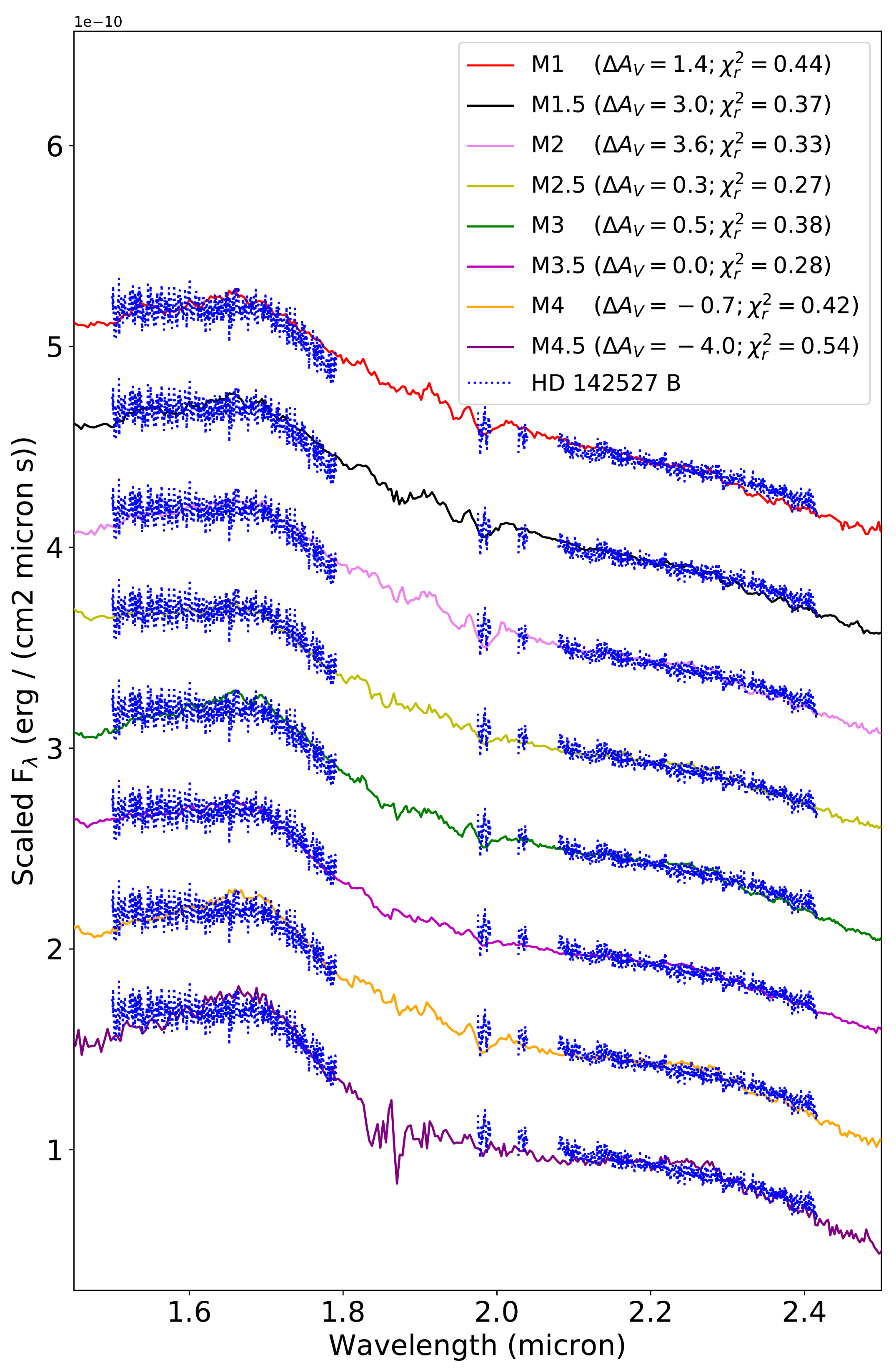}
\caption{ \label{fig_sequence}
Observed $H$+$K$ spectrum of HD~142527~B (\emph{blue dotted curve}) compared to a sequence of young early-M stars built using the speX Prism library \citep{Burgasser2014}. 
The best fit to the spectrum of HD~142527~B is obtained with the M2.5 template. 
Sources are all extracted from young open cluster \object{IC 348} \citep{Luhman2016}: 2MASS J03435953+3215551 (M1), 2MASS J03440216+3219399 (M1.5), 2MASS J03453230+3203150 (M2), 2MASS J03443481+3156552 (M2.5), 2MASS J03430679+3148204 (M3), 2MASS J03435856+3217275 (M3.5), 2MASS J03452021+3217223 (M4) and 2MASS J03450206+3159341 (M4.5).
}
\end{center}
\end{figure}

%
%
A common way to estimate the spectral type of a particular source is to compare its spectrum to a spectral sequence of standard stars.
This was done in the IR for field M- and L-dwarfs \citep[e.g.,][]{Jones1994,Cushing2005}. 
Nevertheless, youth (i.e., low-gravity) is known to significantly affect the shape of the continuum in the NIR \citep[e.g.,][]{Meyer1998,Gorlova2003,Lodieu2008}. 
Considering objects with an estimated age that is similar to our source is critical to avoid bias from the effects of gravity in the spectral-type estimate. 

\citet{Allers2013} proposed a spectral sequence for very-low-gravity objects with spectral types ranging from M5 to L3.
However, the lack of humps at 1.67 and 2.25$\mu$m in the spectrum of HD~142527~B suggests that it is of earlier spectral type than M6. 
Therefore, we built our own sequence of very-low-gravity early- to mid-M stars from the well-studied open cluster \object{IC 348}, located in the nearby Perseus molecular cloud \citep[$\sim 300$~pc;][]{Schlafly2014}.
This open cluster has an estimated age of 2--6 Myr \citep[e.g.,][]{Luhman2003,Bell2013}, which is similar to the estimate for HD~142527~A \citep[$5.0\pm1.5$ Myr old;][]{Mendigutia2014}.
Therefore, similar gravity-sensitive spectral features are expected in our very young M-dwarf sequence and in the spectrum of HD~142527~B.

The sequence is shown in Fig.~\ref{fig_sequence}.
It was built using the SpeX Prism library, which is a compilation of NIR (0.8--2.5 $\mu$m) low-resolution ($\lambda/\Delta \lambda \approx$ 75--120) spectra of red and brown dwarfs obtained with the IRTF SpeX spectrograph \citep{Burgasser2014}.
In particular, the SpeX Prism library contains the spectra of 100 low-mass objects identified to be members of \object{IC 348} and whose spectral types were estimated based on combined optical and NIR spectra \citep{Luhman2016}.
Among the \object{IC 348} objects, we chose a template spectrum for each spectral subtype between M1 and M4.5 with steps of 0.5, based on visual inspection of the quality of the spectra, provided in \citet{Luhman2016}.
Each template spectrum was rescaled to the level of the spectrum of HD~142527~B, and either reddened or unreddened by an amount provided in parenthesis in Fig.~\ref{fig_sequence}, corresponding to the best fit to our SINFONI spectrum.
We chose to make extinction a free parameter given that the SpeX library consists of spectra  that have not been unreddened, and that we are considering very young objects  which can be either more or less self-embedded than HD~142527~B.
The spectrum of HD~142527~B at the bottom of the figure (represented along the M4.5 template) is at the measured flux, while other occurrences of the spectrum are shifted vertically by steps of $5 \times 10^{-11}$ ergs s$^{-1}$ cm$^{-2}$ $\mu$m$^{-1}$.

The best fit is obtained with the M2.5 template ($\chi_r^2 = 0.27$), closely followed by the M3.5 template ($\chi_r^2 = 0.28$).
Slightly poorer fits are obtained with the M1.5, M2.0 and M3.0 templates.
Earlier spectral types than M1.5 are characterized by discrepant slopes at the end of both the $H$- and $K$-bands.
Spectral types later than M4 show significant humps at 1.67 and 2.25$\mu$m, which are the result of the strengthening of the water absorption bands carving the edges of the $H$- and $K$-bands, and are not seen in our spectrum.
This analysis suggests that the spectral type of HD~142527~B is M2.5$\pm$1.0.
We note that the best-fit spectra were obtained without the necessity of applying significant differential extinction  ($\Delta A_V$ = 0.3 and 0.0 for the two best-fit templates), which suggests that the extinction towards HD~142527~B is similar to that towards IC~348.
\citet{Cernis1993} determined that the mean extinction towards IC~348 was $A_V = 2.5 \pm 0.6$ ($A_H \approx 0.4 \pm 0.1$). This constraint will be further discussed in Sect.~\ref{Magnitude}.

\citet{Luhman2003} built a relationship to convert spectral type into effective temperature for very young low-mass objects, based on the assumption that the members of the young GG~Tau quadruple system, spanning K7 to M7.5 in spectral type, were coeval.
This conversion scale is intermediate between those of giants and dwarfs, and yields $T_{\mathrm{eff}} = 3480 \pm 130$K for an M2.5$\pm$1.0 spectral type.
This is in remarkable agreement with the best-fit $T_{\mathrm{eff}}$ obtained with BT-SETTL models ($3500 \pm 100$K; Sect.~\ref{BTSETTL}).

\begin{figure*}[tb]
\begin{center}
\sidecaption
\includegraphics[width=0.7\textwidth]{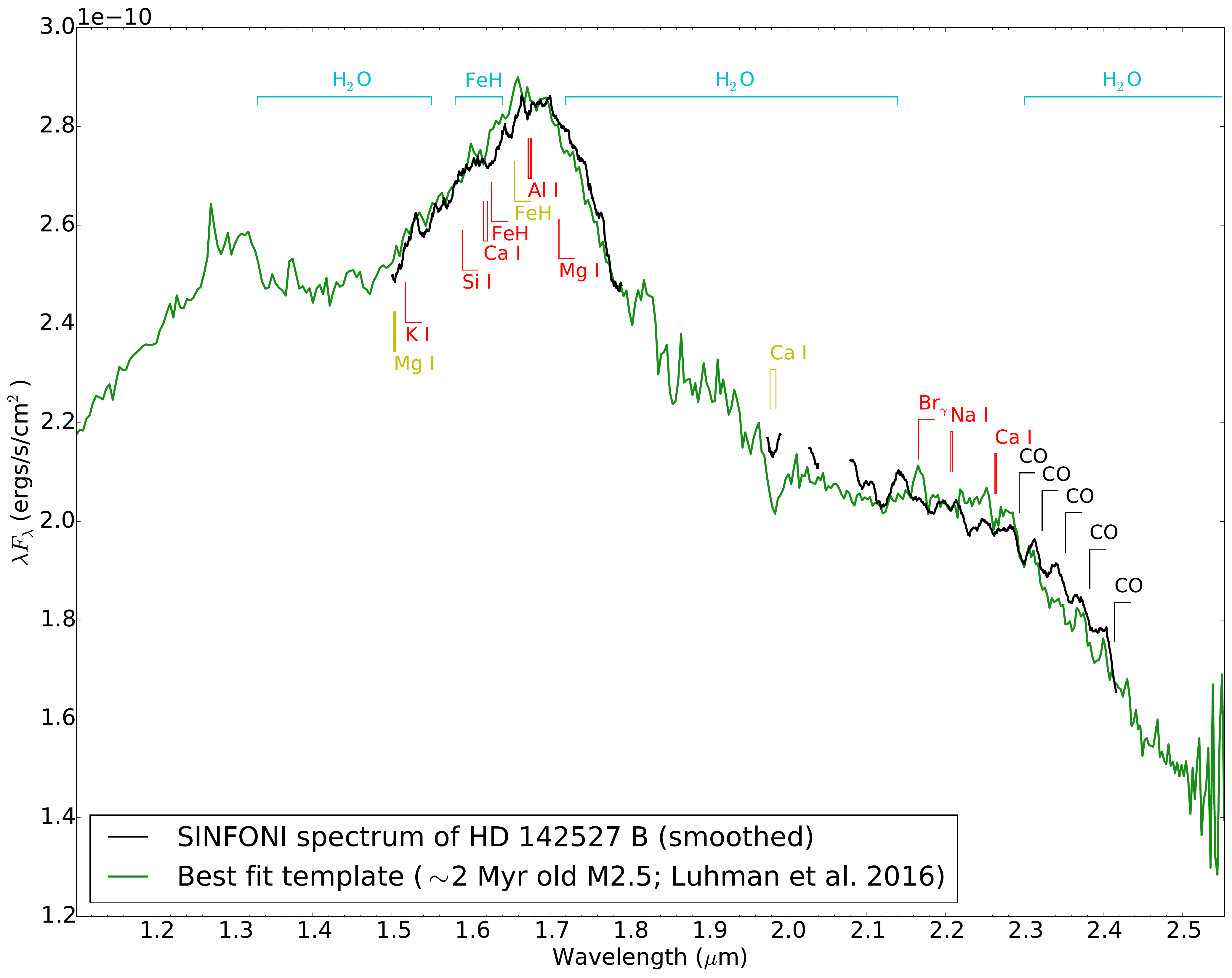}
\caption{ \label{fig_spec_B_speX}
Observed $H$+$K$ spectrum of HD~142527~B (\emph{black curve}) and best-fit SpeX Prism spectrum corresponding to a young M2.5 star (2MASS J03443481+3156552) surrounded by a transition disk in open cluster IC 348 (\emph{green curve}). 
The SINFONI spectrum was Savitzky-Golay filtered to match the resolution of the SpeX spectrum. The template spectrum was rescaled to match the H-band flux of HD~142527~B.
}
\end{center}
\end{figure*}

In addition to the spectral sequence, we considered another test consisting in finding the best-fit template spectrum in the SpeX Prism library.
We considered all objects with a spectral type between M0 and M9 and S/N $>$ 30 in the SpeX library, which totaled 507 objects. 
Due to the lower spectral resolution, we interpolated the SpeX spectra with a spline of degree 3 to reach the same resolution as our SINFONI data.
The fitting process involved two free parameters, corresponding to flux scaling and relative extinction. 
The best-fit was obtained with \object{2MASS J03443481+3156552}; that is, the M2.5 template of the sequence we built in Fig.~\ref{fig_sequence}.
The best-fit template spectrum is shown in Fig. \ref{fig_spec_B_speX}, along with our spectrum of HD~142527~B smoothed to the resolution of the speX spectrum. 
We choose the vertical axis to be in units of $\lambda F_{\lambda}$ in order to better show small changes in spectral slopes.

We notice a remarkable agreement between our smoothed spectrum and the best-fit template in H-band, where most low-resolution absorption features are reproduced.
This instills confidence as to the quality of our extracted spectrum. 
In K-band, the agreement is poorer although most spectral features appear qualitatively reproduced, such as the strong quintuplet of Ca I (1.978--1.986 $\mu$m) and the first overtone of the CO bandhead (2.293--2.414~$\mu$m).
Interestingly, we notice that \object{2MASS J03443481+3156552} was identified as a transition disk \citep{vanderMarel2016a}. 
Its SED\footnote{Available on the Vizier Photometry viewer: \url{http://vizier.u-strasbg.fr/vizier/sed/?-c=&-c.r=10&-c.u=arcsec}.}  indeed shows a strong mid-infrared (MIR) excess compared to the NIR \citep{Dunham2015}.
Therefore slight discrepancies between the spectra of HD~142527~B and the template might be due to differences in the inner part of the respective disk of each star, which is expected to have a stronger influence at $K$ band than at $H$ band.
Finally, we note the presence of the Br$_{\gamma}$ line in the best-fit template spectrum, absent from the spectrum of HD~142527~B. 
This is further discussed in Sect.~\ref{BrG}.

\subsection{Spectral features}\label{SpectralFeatures}

Atomic and molecular features can provide insight into the spectral type and gravity of HD~142527~B.
However, their identification in our spectrum is complicated by the residual speckle noise affecting the flux of the companion in individual channels.
Nevertheless, comparison between our spectrum and BT-SETTL synthetic spectra smoothed at the same spectral resolution (Fig.~\ref{fig_spec_B_modeling}) shows that the residual speckle noise is only slightly larger than the intrinsic noise-like signal corresponding to the forest of absorption lines in the photosphere of M-dwarfs.
This suggests that more information could be extracted from our spectrum.
Therefore, we labeled atomic and molecular lines that are expected in the spectrum of early- to mid-M type stars in Figs.~\ref{fig_spec_B_lines}, \ref{fig_spec_B_modeling}a and \ref{fig_spec_B_speX}. 
Most of these lines are predicted by the best-fit BT-SETTL models (\emph{green curves} in Fig.~\ref{fig_spec_B_modeling}).
In Figs.~\ref{fig_spec_B_modeling}a and \ref{fig_spec_B_speX}, predicted lines that are detected, tentatively detected, and non-detected are shown in \emph{black}, \emph{yellow,} and \emph{red}, respectively.

The most conspicuous absorption feature in our spectrum appears to be the first overtone of CO bandhead ($\Delta \nu =$ 2), with five visible CO transitions, $\nu =$ 2--0, 3--1, 4--2, 5--3 and 6--4, at 2.293, 2.323, 2.352, 2.383 and 2.414~$\mu$m, respectively. 
The first overtone of CO bandhead is a common feature in M- and L-dwarfs, both young and in the field \citep[e.g.,][]{McLean2003,Cushing2005}.
It has also been associated to the inner disk in $>23$\% of T-Tauri stars \citep{Connelley2010}.
Two atomic lines are also tentatively detected in our spectrum: the Mg I triplet (1.502, 1.504 and 1.505 $\mu$m) and the Ca I quintuplet (1.978--1.993 $\mu$m; inset of Fig.~\ref{fig_spec_B_lines}).
The first three lines of the Ca I quintuplet are the strongest absorption features in the $K$-band spectrum of all M-dwarfs, while the Mg I triplet is the strongest absorption feature in $H$ band for M1 to M3 dwarfs \citep[e.g.,][]{Cushing2005}.
For spectral types earlier than M1, Si I (1.589 $\mu$m) becomes the dominant absorption feature in $H$ band, while objects with spectral type later than M4 show a strong K I (1.517 $\mu$m) absorption line.
The non-detection of those two lines, together with the tentative detection of the Mg I triplet and Ca I quintuplet, appears consistent with an estimated M2.5$\pm$1.0 spectral type.

Nonetheless, we note that all transitions of the CO bandhead, the Mg I triplet and Ca I quintuplet are significantly shallower than predicted by the best-fit BT-SETTL synthetic spectrum.
A possible explanation is photospheric line veiling. 
T-Tauri stars are indeed known to show significant line veiling in NIR \citep[e.g.,][]{Greene1996,Folha1999}.
Veiling at optical wavelengths is likely related to hot accretion streams of gas \citep[e.g.,][]{Martin1996} or accretion shocks at the base of magnetospheric accretion columns \citep[e.g.,][]{Calvet1998}.
However, the latter are expected to produce a veiling of only $\sim$0.1 in NIR, insufficient to account for measured veiling ratios $\gtrsim 1.0$ \citep{Folha1999}.
Signicant sources of line veiling in NIR might rather be associated to emission from a hot inner rim and/or low-gravity opacity drop of H$^-$ around 1.67 $\mu$m in the photosphere \citep{Cieza2005,Wing2003,Vacca2011}.
The veiling due to the contribution of a 1700~K inner rim can be seen by comparing the depth of the lines in the green curves of Fig.~\ref{fig_spec_B_modeling}a and b.
However, this effect appears to only partially explain the observed veiling, given that the first three transitions of CO bandhead are still shallower than expected by the best-fit BT-SETTL+hot environment (\emph{green curve} in Fig.~\ref{fig_spec_B_modeling}b).
Moreover, that model predicts that the Ca I doublet (1.616--1.621 $\mu$m), Al I triplet (1.672, 1.676 and 1.677 $\mu$m), Mg I (1.711 $\mu$m) and the Na I doublet (2.206--2.209 $\mu$m) should be detected, while these lines do not appear significant in our spectrum.
The youth of the HD~142527 system (and therefore the low-gravity of the companion) is likely another factor explaining the non-detection of those lines.
A significant $H$-band excess is indeed expected from the reduction of H$^-$ opacity due to the lower density and gravity in the photosphere of young M-stars \citep[e.g.,][]{Wing2003}.
This $H-$band excess was observed in the spectrum of M2.5 T-Tauri star TW~Hya in \citet{Vacca2011}, who suggested that it could be a significant source of line veiling.
Specific examples of previously observed veiled lines in the spectrum of M-type objects in young star-forming associations are the Na I doublet (2.206--2.209 $\mu$m) and K I (1.517 $\mu$m) atomic lines \citep[e.g.,][]{Allers2007,Lodieu2008}.

Finally, we notice another possible line at $\sim$1.65$ \mu$m which, contrarily to other lines we labeled, is not predicted by the best-fit BT-SETTL model (Fig.~\ref{fig_spec_B_modeling}a). 
We tentatively assign it to the FeH absorption line at 1.655$ \mu$m, which is one of two major absorption features in the spectra of mid- to late-M dwarfs \citep[the other one being the FeH bandhead at 1.625 $\mu$m, also labeled in Fig.~\ref{fig_spec_B_modeling}a;][]{Cushing2003,Cushing2005}.
Given that this line is not expected to be significant for spectral types earlier than M5, it could rather be a $\sim 2.5\sigma$--outlier channel in our spectrum; which is statistically possible given that our spectrum is composed of 1313 channels.

\subsection{Flux and magnitude}\label{Magnitude}

Three values for the extinction have been suggested for the \object{HD~142527} system: visible extinctions $A_V=0.60\pm0.05$ (corresponding to $A_H \sim 0.1$) and $A_V=0.80\pm0.06$, both based on the fit of the SED of the primary \citep[][resp.]{Verhoeff2011, Lazareff2017}, and an $H$-band extinction $A_H = 0.3$ based on the NIR colors of the companion \citep{Close2014}. 
In comparison, our best-fit models to the $H$+$K$ spectrum (Sect.~\ref{bestfit_BTSETTL}) suggest values of $A_H = 0.75^{+0.05}_{-0.10}$ in the case where the $H$+$K$ signal is composed of photospheric signal alone, and $A_H = 0.0$--$0.2$ if a hot circum-secondary environment is present.
A value of $A_H = 0.75^{+0.05}_{-0.10}$ could suggest that the companion is self-embedded, and suffers much more extinction than the primary.
On the contrary, $A_H = 0.0$--$0.2$ is consistent with the estimate of the extinction towards the primary. 
 The best-fit template spectra (Sect.~\ref{realspectra}) suggest that a similar extinction is affecting observations towards HD~142527~B and members of the IC~348 cloud. In the latter case, \citet{Cernis1993} inferred $A_H \approx 0.4 \pm 0.1$. Given that this value lies roughly in the middle of the two best-fit values of $A_H$ found with BT-SETTL models, it does not constrain whether a hot circum-secondary environment is contributing or not to the $H$+$K$ spectrum.
However, among the best-fit BT-SETTL+environment models (consisting in ranges of values in $R_B$, $A_H$, $R_d$ and $T_d$; Table~\ref{tab:best_fit_params}), the best-fit solution with $A_H = 0.2$ leads to a better agreement with the photometric measurement at R-band, which is the most sensitive to extinction (Fig.~\ref{fig_spec_B_modeling}d).
Best-fit solutions with $A_H < 0.2$ lead to $\gtrsim2\sigma$--discrepancy with the R-band photometric point.
Therefore, for the rest of the analysis we favor the best-fit solution with $A_H = 0.2$ (for the case where a hot circum-secondary environment is present), which is also in better agreement with $A_H \approx 0.4 \pm 0.1$ inferred for the IC~348 cloud.

Several indications hint towards the presence of a circum-secondary disk, such as (i) a better-quality fit obtained with the inclusion of a hot circum-secondary environment, (ii) the best-fit template obtained with a young M2.5 star surrounded by a transition disk, and (iii) the detection of $H_{\alpha}$ emission suggesting on-going mass accretion \citep{Close2014}.
Nevertheless, we cannot rule out the possibility that the putative circum-secondary disk has a negligible contribution in $H$+$K$ and that the poorer fit with the synthetic spectra alone is due to the inappropriateness of the BT-SETTL models at such young age. 
Therefore, we continue to consider the two possibilities in the remainder of the analysis; that is, the $H$+$K$ spectrum is composed of either photospheric signal alone (case I), or the combination of photospheric+hot circum-secondary environment contributions (case II).
Table \ref{tab:Magnitude_B} compiles the measured flux, measured apparent magnitude and absolute de-reddened magnitude of the companion in each of those two cases.
The absolute de-reddened magnitudes are computed using the corresponding best-fit extinction value: $A_H=0.75$ and $0.2$ without and with a circum-secondary environment, respectively.
In case II, 
only the contribution from the photosphere (\emph{cyan curve} in Fig.~\ref{fig_spec_B_modeling}d) is provided in Table~\ref{tab:Magnitude_B}. 
Making this distinction (i.e., subtracting the contribution from the disk) is necessary for an appropriate placement of the companion in HR diagrams (Sect.~\ref{CMD}).
The quoted uncertainties in flux and apparent magnitude in Table~\ref{tab:Magnitude_B} consider a conservative 10\% relative flux uncertainty (0.11~mag) and, for case II, an additional error added in quadrature representing the uncertainty on the radius of the companion (Table~\ref{tab:best_fit_params}).
The uncertainty in absolute de-reddened magnitude also includes the uncertainty on $A_H$ (Table~\ref{tab:best_fit_params}).

\begin{table} [t]
\begin{center}
\caption{Flux and magnitude of HD 142527 B in $H$ and $K$ band.}
\label{tab:Magnitude_B}
\begin{tabular}{lcccc}
\hline
Band & Case & Flux & Apparent & Absolute \\
 & & [mJy] & magnitude & magnitude\\
\hline
$H$ & I & $151 \pm 15$ & 9.63 $\pm$ 0.11 & $2.92^{+0.15}_{-0.12}$ \\
 & II & $116 \pm 10$ & 9.92 $\pm$ 0.16 & $3.75^{+0.25}_{-0.16}$ \\
$K$ & I & $147 \pm 15$ & 9.14 $\pm$ 0.11 & $2.71^{+0.15}_{-0.12}$ \\
  & II & $98 \pm 8$ & 9.59 $\pm$ 0.16 & $3.50^{+0.25}_{-0.16}$ \\
\end{tabular}
\end{center}
Notes:
Case I considers that all the observed $H$+$K$ flux comes from the photosphere of the companion.
Case II considers the contribution from the photosphere alone in the best-fit BT-SETTL+hot environment model to the $H$+$K$ spectrum.
The absolute de-reddened magnitudes are computed using $A_H = 0.75$ and $0.2$ for cases I and II, respectively.
\end{table}

The filter transmission curves of CONICA for $H$ and $K$ are overlapping with wavelength areas that are significantly affected by telluric lines, and are therefore less reliable.
For this reason, the flux and magnitude of the companion derived from our SINFONI spectrum were computed using the $H$ and $K$ filter transmission curves of 2MASS. 
We do not expect a significant change in the estimation of the magnitude, in particular when compared to other sources of uncertainties on the flux of the companion. 

\subsection{Mass and age}\label{CMD}

\begin{figure}[t]
\begin{center}
\includegraphics[width=0.48\textwidth]{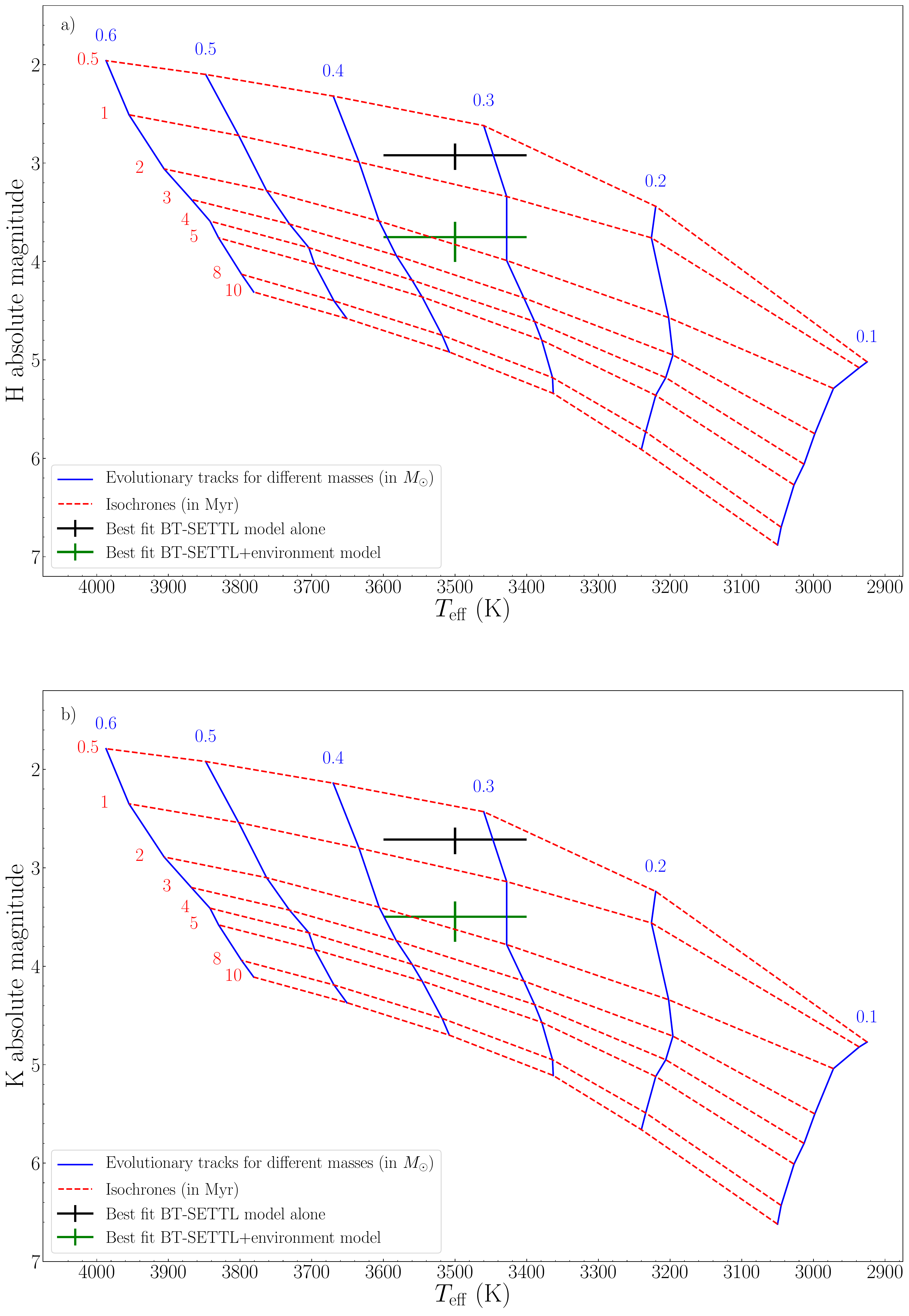} 
\caption{ \label{fig:CMDs}
HR diagrams in 
$H$- (a) and $K$-band (b) showing stellar tracks for different masses according \citet{Baraffe2015} evolutionary models.
Isochrones (\emph{red-dashed lines}) range from 0.5 to 10 Myr.
The evolutionary paths (\emph{blue lines}) are considered for stellar masses ranging from 0.1 to 0.6~$M_{\sun}$, per steps of 0.1$M_{\sun}$.
The \emph{black} and \emph{green error bars} correspond to the best-fit $T_{\mathrm{eff}}$ and dereddened absolute magnitude of the companion in case all the observed $H$+$K$ signal is photospheric (case I) or made of a combination of photosphere+hot environment (case II).
}
\end{center}
\end{figure}

The standard procedure to estimate stellar mass and age consists in comparing the star location with evolutionary tracks in an HR diagram \citep[e.g.,][]{Siess2000, Bressan2012}.
Here, we choose to use the \citet{Baraffe2015}\footnote{Available at \url{http://perso.ens-lyon.fr/france.allard/}.} evolutionary models, which assume a solar metallicity with the revised heavy element fraction by mass $Z = 0.0153$ from \citet{Caffau2011}. 
Figure~\ref{fig:CMDs}a and b shows the $H$- and $K$-band HR diagram comparing the location of HD~142527~B with the evolutionary tracks of young stellar objects ranging from 0.1 to 0.6~$M_{\sun}$.
Isochrones were considered between 0.5 and 10~Myr.
The $T_{\mathrm{eff}}$ of HD~142527~B was found to be $3500 \pm 100$K based on best-fit models using BT-SETTL synthetic spectra (Sect.~\ref{BTSETTL}). 
This value is consistent with the $T_{\mathrm{eff}}$ corresponding to the best-fit spectral type (M2.5$\pm$1.0) inferred from young spectral templates: $3480 \pm 130$K (Sect.~\ref{realspectra}).
Therefore we consider a single value of $T_{\mathrm{eff}} = 3500 \pm 100$K in the HR diagrams.
However, we consider two values for the absolute de-reddened magnitude of the companion, corresponding to the cases of the absence or presence of a hot circum-secondary environment (cases I and II in Table~\ref{tab:Magnitude_B}).

In each considered case, we notice that the location of the companion leads to consistent mass and age estimates in the $H$- and $K$-band HR diagrams.
If all the observed $H$+$K$ signal is photospheric (case I), the estimated mass and age of HD~142527~B are $0.33\pm0.05 M_{\sun}$ and $0.75\pm0.25$ Myr, respectively.
Considering the best-fit photospheric+hot environment model (case II), the mass and age of the companion are $0.34\pm0.06 M_{\sun}$ and $1.8_{-0.5}^{+1.20}$ Myr, respectively.
We note that in the first case, the very young age is compatible with the significant extinction inferred from the fit to BT-SETTL models ($A_H = 0.75$) conveying the fact that the companion is still embedded in its birth environment.
Despite the different age estimates, both cases lead to very similar mass estimates, which is due to the verticality of the Hayashi tracks in the $H$- and $K$-band HR diagrams in the considered range of stellar mass and age.

\subsection{Radius}\label{radius}

For the derived mass and age of $0.33\pm0.05~M_{\sun}$ and $0.75\pm0.25$ Myr (case I), the \citet{Baraffe2015} evolutionary models predict a stellar radius of $1.96\pm0.10~R_{\sun}$. 
This is consistent with the best-fit stellar radius obtained independently by calibrating in flux the BT-SETTL model to the $H$+$K$ spectrum ($2.08 \pm 0.18~R_{\sun}$).
In case II, the \citet{Baraffe2015} evolutionary models predict a stellar radius of $1.37 \pm 0.05~R_{\sun}$ for a mass and age of $0.34\pm0.06 M_{\sun}$ and $1.8_{-0.5}^{+1.20}\pm$ Myr. 
This is also consistent with the stellar radius value of 1.42 $R_{\sun}$ found independently from the flux adjustment of the best-fit BT-SETTL+hot circum-secondary environment models to our spectrum.
These values of stellar radius indicate that the M2.5$\pm$1.0 companion is still gravitationally contracting, as expected in such a young system. 

It is also possible to estimate the expected ratio of stellar radii between the primary and the companion:
\begin{equation}
\frac{R_A}{R_B} = \sqrt{\frac{F_A T_{\mathrm{eff},B}^4}{F_B T_{\mathrm{eff},A}^4}},
\end{equation}
 based on the bolometric flux of the best-fit BT-SETTL models for HD~142527~A ($F_A$) and HD~142527~B ($F_B$), and the best-fit effective temperatures for the primary ($T_{\mathrm{eff},A}$) and the companion ($T_{\mathrm{eff},B}$).
Considering $T_{\mathrm{eff},A} = 6500$~K, $T_{\mathrm{eff},B} = 3500$~K,  the model shown in \emph{black} in Fig.~\ref{fig_spec_A} to compute $F_A$, and the model shown in green (resp. cyan) in Fig.~\ref{fig_spec_B_modeling}c (resp.~\ref{fig_spec_B_modeling}d) to calculate $F_B$ leads to $\frac{R_A}{R_B} = 1.52$ (resp. 2.25) in case I (resp. II).
These ratios are consistent with $\frac{R_A}{R_B} = 1.63 \pm 0.13$ (resp. $2.34 \pm 0.17$) considering the respective best-fit BT-SETTL models leading to $R_A = 3.2\pm0.2 R_{\sun}$ (Table~\ref{tab:HD142527A}) and $R_B = 1.96\pm0.10 R_{\sun}$ (resp. $1.37 \pm 0.05 R_{\sun}$) in case I (resp. II).

\subsection{Surface gravity}\label{SurfaceGravity}

For the combinations of mass and age derived in both cases I and II, it is noteworthy that the \citet{Baraffe2015} evolutionary models predict lower values of $\log(g)$ ($3.38\pm0.14$ and $3.70\pm0.18$) than inferred from the best-fit BT-SETTL models ($4.5_{-0.5}$ for both cases I and II; Table~\ref{tab:best_fit_params}).
This discrepancy could be due to a significant magnetic field, which is expected to be present for young M-type T-Tauri stars \citep[e.g.,][]{Johns-Krull1999,Johns-Krull2007}.
Several studies have indeed shown that not including a strong ($\sim$ kG) magnetic field can lead to $\sim$0.5-1.0 dex difference in the estimated $\log(g)$ due to an improper account of line broadening in the synthetic spectrum \citep[e.g.,][Flores et al. in prep.]{Doppmann2003,Sokal2018}.
However, these conclusions are based on the difference in line widths in normalized spectra and, to our knowledge, the effect on the continuum shape (which likely dominates our best-fit solutions) has not yet been studied.
Nonetheless, given the young age of the system and the fact that BT-SETTL models do not take into account the effect of magnetic fields, we suspect the surface gravity value predicted by evolutionary models ($\log(g)\sim$3.38 and 3.70 in cases I and II, resp.) to be more representative of HD~142527~B, and speculate that a substantial magnetic field \citep[$\sim$1-2 kG, if similar to other M-type T-Tauri stars;][]{Johns-Krull1999,Johns-Krull2007} might explain the $\sim$0.5-1.0 dex discrepancy in $\log(g)$. 

\subsection{Mass accretion rate} \label{BrG}

\citet{Meyer1997} suggested that the observed dereddened $H - K$ color of T-Tauri stars can be related to their mass accretion rate.
The absolute $H$- and $K$-magnitudes of HD~142527~B yield $H - K \approx 0.2 \pm 0.2$, 
which leads to a weak constraint of $\dot{M_B} \lesssim 10^{-8} M_{\sun} \mathrm{yr}^{-1}$ \citep[see Fig.~5 in][]{Meyer1997}.

\citet{Calvet1991} showed that for T-Tauri stars with effective temperature lower than 4000~K, the CO bandhead associated to the disk is expected to be seen in emission for mass accretion rates $\lesssim 10^{-8} M_{\sun} \mathrm{yr}^{-1}$,  which would therefore veil the photospheric CO bandhead. 
In Fig.~\ref{fig_spec_B_modeling}a (resp. \ref{fig_spec_B_modeling}b), comparison between the observed depth of the CO bandhead and the expected depth of the CO bands based on the BT-SETTL (resp.~BT-SETTL+hot environment) model
indeed suggests significant (resp. moderate) veiling. 
Therefore, similarly to the $H - K$ color, the observed veiling of CO also constrains the mass accretion rate of the companion to be $\dot{M_B} \lesssim 10^{-8} M_{\sun} \mathrm{yr}^{-1}$. 

\citet{Connelley2010} noted that the presence of veiled photospheric CO was often accompanied by Br$_{\gamma}$ emission, a known tracer of mass accretion \citep[e.g.,][]{Muzerolle1998}.
Nevertheless, our spectrum of HD~142527~B does not show any significant departure from continuum emission around the expected wavelength of the Br$_{\gamma}$ line. 
Considering an optimistic 5\% relative uncertainty (Sect. \ref{errors} and Appendix~\ref{app:uncertainties}) and a $-1$~nm equivalent width \citep[the rough value for the Br$_{\gamma}$ line of the primary;][]{Mendigutia2014}, the 3$\sigma$--upper limit on the line flux is $\sim 1.5 \times 10^{-14}$~ergs~s$^{-1}$~cm$^2$. 
The Br$_{\gamma}$ luminosity is given by:
\begin{equation}
L(\mathrm{Br}_{\gamma}) =  4\pi D^2 \times F(\mathrm{Br}_{\gamma}),
\end{equation}
where $D$ is the distance, assumed to be 156~pc, and $F(\mathrm{Br}_{\gamma})$ is the total dereddened Br$_{\gamma}$ line flux. 
We find $\log (L(\mathrm{Br}_{\gamma}) / L_{\sun}) \lesssim -5.0$. 
The accretion luminosity is related to the Br$_{\gamma}$ line luminosity with the following empirical expression, applicable down to low-mass T-Tauri stars \citep{Calvet2004}:
\begin{equation}
\log (L_{\mathrm{acc}}/ L_{\sun}) = 0.9 \times (\log L(\mathrm{Br}_{\gamma}) / L_{\sun} + 4) -0.7.
\end{equation}
The final accretion luminosity of \object{HD~142527~B} is found to be $\lesssim 0.026 L_{\sun}$. 
This is compatible with the 0.013 $L_{\sun}$ estimated from the H$_{\alpha}$ line detection presented in \citet{Close2014}.
We therefore conclude that residual speckle noise alone can account for the non-detection of Br$_{\gamma}$ emission.

Using the accretion luminosity to mass accretion rate relationship $\dot{M_B} = 1.25 L_{\mathrm{acc}} R_B/G M_B$ \citep{Gullbring1998},
with the value of $L_{\mathrm{acc}}$ found in \citet{Close2014}, but with the new values of mass and radius derived in this work yields $\dot{M_B} \sim 5.8 \times 10^{-9} M_{\sun} \mathrm{yr}^{-1}$ (resp. $\dot{M_B}  \sim 4.1 \times 10^{-9} M_{\sun} \mathrm{yr}^{-1}$) for the companion in case I (resp. case II). 
This corresponds to about 3\% (resp. 2\%) of the mass accretion rate on the primary, in case I (resp.~II). 
We note that these estimates are consistent with the constraints based on the appearance of the CO bandhead and the dereddened $H - K$ color ($\dot{M_B} \lesssim 10^{-8} M_{\sun} \mathrm{yr}^{-1}$).

\section{Discussion}\label{discu}

\subsection{Comparison with previous works}\label{SEDfitting}

The nature and orbit of the companion have recently been investigated by \citet{Lacour2016}.
They suggest that a low-mass star with an effective temperature of 3000~K and a surface gravity $\log(g) = 3.5$ surrounded by a 1700~K circum-secondary environment can account for the whole SED of the companion, considering a visible extinction $A_V$=0.6. 
We compare our $H$+$K$ spectrum with the model suggested in \citet{Lacour2016} in Fig.~\ref{fig_spec_B_modeling}b, 
and note a disagreement between the shape of our spectrum and their model. 
We also notice a significant discrepancy between our measured $H$- and $K$-band fluxes and the ones measured with SAM \citep{Biller2012,Lacour2016}.
While our measurements are within $2\sigma$ compared to the fluxes measured in \citet{Biller2012}, they correspond to $>3\sigma$--differences with the measurements in \citet{Lacour2016}.
In particular, both the $H$- and $K$-band photometric measurements are about twice fainter than our spectrum.
Both our tests (Appendix~\ref{app:uncertainties}) and the consistency between our estimated radius and the radius predicted by evolutionary models suggest that the absolute flux calibration of our spectrum is of good quality.
Even if the uncertainty in absolute flux of our spectrum was slightly larger than 15\% (Appendix~\ref{app:uncertainties}), it would still be incompatible with the 2016 NACO-SAM $H$- and $K$-band measurements.
Investigating the possible origins of such a difference with the SAM measurements would require an in-depth study which is beyond the scope of this paper.
Nevertheless, we note that new measurements with extreme-AO instruments could provide new insight regarding this discrepancy.

Not surprisingly, the significantly fainter SAM photometric measurements led to a much smaller estimated mass for the companion: 0.1 $M_{\sun}$, corresponding to the lower-mass end of the initial estimate of 0.1--0.4~$M_{\sun}$ \citep{Biller2012}.
On the contrary, our new analysis leads to a mass estimate of $\sim 0.35 M_{\sun}$, thus closer to the upper limit of that range.
In the case that all the $H$+$K$ observed flux is photospheric, our estimated age of HD~142527~B ($0.75 \pm 0.25$ Myr) is significantly younger than the age derived for the primary \citep[$5.0 \pm 1.5$ Myr;][]{Mendigutia2014,Lacour2016}.
This would suggest that the companion formed after the primary, and its extremely young age would be consistent with the significant extinction to be applied to BT-SETTL synthetic spectra to fit the observed spectrum ($A_H = 0.75^{+0.05}_{-0.10}$), conveying that the companion is still embedded in its birth environment.
On the contrary, the best-fit model including a hot circum-secondary disk leads to an estimated age of $1.8^{+1.2}_{-0.5}$ Myr, hence only slightly discrepant with the age of the primary.

Based on the measured $H_{\alpha}$ luminosity of the companion,
\citet{Close2014} found a mass accretion rate of $\sim5.9 \times 10^{-10}$~M$_{\sun}$ yr$^{-1}$. 
Nevertheless, at that time the mass and radius of the companion were poorly constrained.
Our new estimate of the mass accretion rate based on the values of mass and radius inferred in this work is almost an order of magnitude larger ($4.1$--$5.8 \times 10^{-9}$ M$_{\sun}$ yr$^{-1}$). 
Given that the new mass ratio between the companion and the primary is $q \sim 1/6$, a mass accretion rate estimate which is 2--3\% that of the primary does not appear unreasonable.

\subsection{Circumbinary or transition disk?}\label{circumbinary}

The wealth of features found in the disk of  \object{HD~142527}, including a large gap, a horseshoe-shaped millimeter continuum and several spiral arms, have been considered as signposts of planet formation \citep[e.g.,][]{Ohashi2008,Casassus2013}. 
Similarly to HD~142527, the majority of large transition disks (with gaps larger than 20 au) show high accretion rates \citep[e.g.,][]{Manara2014}, 
further suggesting that the origin of the gap is  
not related to photo-evaporation, but possibly to embedded companions \citep[see][and references therein]{Owen2016}.

Our analysis of the spectrum of HD~142527~B suggests that the mass ratio with respect to the primary is substantial ($q\sim1/6$).
 Our new mass estimate was recently considered to inject the binary companion in hydro-dynamical simulations which showed that most of the observed characteristics of the disk could be explained by the interaction of the binary with the disk \citep{Price2018}.
One is therefore tempted to label HD~142527 a circumbinary disk, even if the term is more often used to refer to equal-mass binaries, as in GG~Tau~A and CoKu Tau~4 \citep[e.g.][]{Ireland2008,Beck2012}. 
However, we note that for a given mass ratio between the primary and companion, disks around lower-mass primaries could show similar features as \emph{circumbinary disks} but still be called \emph{transition disks} due to a substellar companion.
Therefore, the distinction between circumbinary disks and (at least a fraction of) transition disks may be purely semantic, as similar physical processes could be at the origin of the observed disk morphologies.
 
The fact that a $\sim 0.35 M_{\sun}$ stellar companion evaded detection until recently causes us to question the prevalence of gapped disks where a binary or massive companion has not been detected yet.
Could less massive companions (with e.g., $q\sim0.1$--$0.01$) be at the origin of similar disk morphologies as HD~142527 with less extreme characteristics (smaller gap, less pronounced dust trap)?
Several surveys have investigated the presence of massive companions in transition disks, using two main methods: radial velocities and SAM \citep[e.g.,][]{Ghez1993, Simon1995, Pott2010}.
Recent results from radial velocity surveys, probing the innermost regions of transition disks, suggest that the latter do not host a significantly larger number of spectroscopic binaries than field stars \citep[$\sim 4$\%; ][]{Kohn2016}.
However, the most recent survey making use of SAM led to the conclusion that up to $\sim$40\% of TDs could in fact be circumbinary disks \citep{RuizRodriguez2016}.
Nonetheless, these studies did not make distinctions between, for example, millimeter-bright and millimeter-faint disks within the larger class of transition disks.
Furthermore, most of these surveys have focused on transition disks in low-mass star forming regions (mostly K and M stars), and thus did not include any Herbig Ae/Be/Fe star. 
For Herbig stars, detections of companions down to the low-mass end of the stellar regime still remains very challenging \citep[e.g.,][]{Duchene2013}. 
Here the companion is $\sim 0.35 M_{\sun}$ and lies at 14 au from the central star, but was only detected in 2012. 
The companion would likely have been missed if it either lied slightly closer to the primary or was slightly less massive. 
Large gaps in millimeter-bright Herbig disks could therefore possibly be carved by close companions (either stellar or massive planetary) falling just below the detection limits in terms of either inner working angle (for direct high-contrast imaging) or achieved contrast (for SAM).

For those gapped disks harboring a binary or massive substellar companion, recent simulations point towards a much more complex disk evolution than previously expected, involving a misalignment between the circumbinary disk and either the binary orbital plane \citep{Martin2017} or the inner disk plane \citep{Owen2017}.
Disks with complex morphologies (inner cavity, spiral arms, asymmetric millimeter-dust distributions, warp) are therefore prime targets to search for companions.
As a matter of fact, high-contrast thermal-IR imaging has recently revealed a candidate companion in the gapped disk of Herbig Ae star MWC~758  \citep{Reggiani2018}.
The companion is thus-far poorly characterized, but is found in a situation that is reminiscent of HD~142527~B: inside  an eccentric sub-millimeter cavity and internal to both multiple spiral arms and dust traps in the outer disk \citep[e.g.,][]{Benisty2015,Dong2018}.
Detection and characterization of more companions in those disks and subsequent hydrodynamical modeling could answer the question of whether the observed disk morphology can be traced back to a common origin: dynamical interaction with an embedded companion.



\section{Summary}\label{ccl}

In this paper, we confirm the spectral type of HD~142527~A to be F6$\pm$0.5III-V, based on our VLT/SINFONI spectrum and a re-analysis of VLT/X-SHOOTER data.
ADI post-processing algorithms were used on the $H$+$K$ spectral channels of our VLT/SINFONI observations of the HD~142527 system to obtain the first medium-resolution spectrum of a low-mass companion at less than 0\farcs1~from its central star.
We used the NEGFC technique to derive the position and contrast of \object{HD~142527~B} in each channel.
A careful estimation of the uncertainties related to speckle noise allowed us to derive robust error bars on our spectrum, typically of order 5--10\% relative uncertainty in flux.
Our final astrometry is the following: $r = 78.0 \pm 1.8$~mas and PA~$= 119.1 \pm 1.0\degr$ on May 10, 2014.
The position is in agreement with previous detections of the companion at similar epoch, however the flux found in $H$ and $K$ bands is not in agreement with previous VLT/NACO sparse aperture masking photometric measurements of the companion.

We compared our spectrum of the companion to BT-SETTL synthetic spectra covering a large range of effective temperature and surface gravity, and considered several additional parameters including the stellar radius, the extinction, and the physical extension and temperature of a putative hot circum-secondary inner rim.
Fitting of the $H$+$K$ spectrum with BT-SETTL synthetic spectra, in the absence of hot circum-secondary material contribution, led to a best-fit effective temperature of 3500 $\pm$ 100~K, a surface gravity $\log(g)=4.5_{-0.5}$, a stellar radius of $2.08 \pm 0.18 R_{\sun}$ and an $H$-band extinction $A_H=0.75^{+0.05}_{-0.10}$ mag.
However, the inclusion of two additional free parameters accounting for a hot circum-secondary environment led to a slightly better fit of the observed spectrum.
The best-fit effective temperature and surface gravity were found to be the same: 3500$\pm$100~K and  $\log(g)=4.5_{-0.5}$.
However, we noticed some degeneracy between the stellar radius, $H$-band extinction, temperature and radius of the circum-secondary environment (assuming a disk shape for simplification), whose best-fit values range in: $R_B \in [1.09, 1.55] R_{\sun}$, $A_H \in [0.0, 0.2]$,  $T_d \in [1500, 1700]$~K and $R_d \in [3.0, 13.9] R_{\sun}$.
Favoring the best-fit model with an $H$-band extinction $A_H = 0.2$ leads to an estimated stellar radius, inner rim temperature, and physical extension corresponding to $R_B = 1.42 R_{\sun}$, $T_d = 1700$K and the area of a disk of $11.8 R_{\sun}$, respectively.
This choice is motivated by the better agreement in $R$-band with the measured flux in \citet{Close2014}.

We also compared our spectrum to a spectral sequence of young low-mass objects taken from the SpeX Prism library. 
The best fit was obtained with the spectrum of a young M2.5 star surrounded by a transition disk, which remarkably reproduced all the observed spectral features in the $H$ band and qualitatively reproduced the $K$-band spectrum of HD~142527~B. 
Using the spectral type to effective temperature conversion scale established in \citet{Luhman2003}, we found that an M2.5$\pm$1.0 spectral type corresponded to an effective temperature of 3480$\pm$130~K, which is in good agreement with the effective temperature inferred from the fit to BT-SETTL models.

We subsequently compared the best-fit effective temperature and the absolute de-reddened magnitudes of HD~142527~B with isochrones of low-mass stars synthesized from the \citet{Baraffe2015} evolutionary models in $H$- and $K$-band HR diagrams.
This allowed us to derive the following estimates for the mass, age, and radius of the companion: 0.34 $\pm$ 0.06~$M_{\sun}$ (resp. 0.33 $\pm$ 0.05~$M_{\sun}$), $1.8^{+1.2}_{-0.5}$~Myr (resp. $0.75 \pm 0.25$~Myr) and $1.37 \pm 0.05 R_{\sun}$ (resp. 1.96 $\pm$ 0.10~$R_{\sun}$), in the case of the presence (resp. absence) of a hot circum-secondary environment contributing to the observed $H$+$K$ band spectrum.
We find that the stellar radius estimates obtained from the \citet{Baraffe2015} evolutionary models are consistent with the best-fit stellar radius obtained independently by calibrating in flux the BT-SETTL models to the $H$+$K$ spectrum. 
We argue that the corresponding surface gravity value $\log(g) = 3.70\pm0.18$ (resp. 3.38$\pm$0.14) in the presence (resp. absence) of hot circum-secondary environment is likely to be more representative of the true surface gravity of HD~142527~B than the value inferred from the best-fit BT-SETTL models ($\log(g)=4.5_{-0.5}$). 
This is justified by the young age of the system and the fact that the effect of a strong magnetic field, typical of M-type T-Tauri stars \citep[e.g.,][]{Johns-Krull2007}, is not taken into account in BT-SETTL synthetic spectra. 

The age estimate of the companion in the case of non-significant disk contribution in $H$ and $K$ bands ($0.75 \pm 0.25$~Myr) appears significantly younger than the value of 5.0 $\pm$ 1.5~Myr derived for the primary \citep{Mendigutia2014}, suggesting that the companion formed after the primary. 
However, the age estimate found in the case of the presence of hot circum-secondary environment is only slightly discrepant with the age of the primary. 
The new mass estimate that we derived (0.34 $\pm$ 0.06~$M_{\sun}$ considering both cases discussed above) is significantly larger than the previous estimate based on SED fitting \citep[0.1~$M_{\sun}$;][]{Lacour2016}. 
This suggests that the impact of the companion on the disk morphology could be more significant than previously expected.
New hydro-dynamical simulations taking into account the mass derived in this work have recently shown that the peculiar morphology of the disk of HD~142527 (large gap, spiral arms, dust trap) could be reproduced by the companion alone considering an inclined and eccentric orbit \citep{Price2018}.

Based on the non-detection of Br$_{\gamma}$ line emission for the companion, we estimated an upper limit on the accretion luminosity compatible with the value computed from the H$_{\alpha}$ detection \citep{Close2014}.
We provide a new estimate of the mass accretion rate of $\sim $4.1--5.8$ \times 10^{-9}$~M$_{\sun}$ yr$^{-1}$ 
based on the H$_{\alpha}$ accretion luminosity and using the new values of mass and radius derived for the companion.
This is roughly 2--3\% of the mass accretion rate on the primary.

We note that the new spectral type of M2.5 makes HD~142527~B a twin of the well known TW~Hya T-Tauri star, for which \citet{Vacca2011} estimated an effective temperature and mass of $3400 \pm 200$~K and $0.4 \pm 0.1~M_{\odot}$, respectively.
Although consistent with our estimates for HD~142527~B, the slightly different values they quote is likely due to the fact they used temperature and mass calibrations based on field M-dwarfs \citep{Leggett1996,Reid2005}. 
The revised age of TW~Hya \citep[$\sim$ 3~Myr;][]{Vacca2011} is also similar to the value we inferred for HD~142527~B.
The new spectral analysis of TW~Hya presented in \citet{Sokal2018} suggests the presence of a strong magnetic field, which is also the interpretation we favor for HD~142527~B to account for the derived values of $\log(g)$.
The only major difference in the $H$+$K$ spectra of both objects is the presence of a significant Br$_{\gamma}$ emission line for TW~Hya \citep{Vacca2011}.

Spectral characterization with extreme-AO might refine the parameters of the companion. 
In particular, a spectrum in $J$ band could enable a better estimate of the extinction towards the companion and a better diagnostic of the gravity of HD~142527~B \citep[see e.g.,][]{Allers2013}. 
New observations in thermal-IR might also confirm the presence of a circum-secondary disk.
Follow-up observations may therefore enable one to discriminate between the two possibilities presented in this paper; that is, whether a putative circum-secondary disk makes a significant contribution to the $H$+$K$ spectrum or not, which would in turn lead to a better estimate of the age of the companion.

From a broader point of view, we demonstrate the efficiency of VLT/SINFONI used in pupil-tracking for the spectral characterization of faint low-mass companions very close to their parent star.
While the companion presented in this paper is in the red dwarf regime, using this mode with a fainter primary star could enable the spectral characterization of close-in substellar companions.

\begin{acknowledgements}

We acknowledge the anonymous referee for his/her constructive comments.
We are grateful to the VLT/SINFONI telescope staff (Kora Muzic and Ivan Aranda) for their precious help 
during the observing run.
We thank Micka\"el Bonnefoy, Jackie Faherty, Gregory Herczeg, Sylvestre Lacour and Gerrit van der Plas for useful discussions. 
We are also thankful to Ignacio Mendigut\'ia for sharing the X-SHOOTER spectrum of HD~142527.
VC acknowledges support
from CONICYT through CONICYT-PCHA/Doctorado Nacional/2016-21161112.
VC and SC acknowledge support by the Millennium Science 
Initiative (Chilean Ministry of Economy), through grant RC130007. VC, OA, CGG and OW acknowledge support from the European Research Council under the European Union's Seventh Framework Program 
(ERC Grant Agreement n. 337569) and from the French Community of Belgium through an ARC grant for Concerted Research Action.
AZ acknowledges support from the CONICYT + PAI/ Convocatoria nacional subvenci\'on a la instalaci\'on en la academia, convocatoria 2017 + Folio PAI77170087
This research has benefitted from the SpeX Prism Library (and/or SpeX Prism Library Analysis Toolkit), maintained by Adam Burgasser at \url{http://www.browndwarfs.org/spexprism}.

\end{acknowledgements}

\bibliographystyle{aa} 
\bibliography{HD142527B_accepted}

\begin{appendix}

\section{Spectral classification of HD~142527~A} \label{app:HD142527A}

The spectral type of HD~142527~A was first estimated to be F6-F7III \citep{Houk1978, vandenAncker1998}. 
More recently, based on a VLT/X-SHOOTER spectrum and on an updated distance estimate, \citet{Mendigutia2014} concluded more conservatively that the spectral type must be comprised between F3 and F8.
In view of the large uncertainty, we re-analyzed the optical part of their X-SHOOTER spectrum (program 084.C-0952). 
The data were loaded from the ESO archive, reduced using the ESO X-SHOOTER pipeline v 2.8.4. \citep{Modigliani2010}, and corrected for telluric absorption using \emph{molecfit} \citep{Smette2015, Kausch2015}. 
The spectral classification was then obtained by comparing directly the X-SHOOTER spectrum with the spectra of all F stars in the {\it X-SHOOTER Spectral Library DR1} \citep[XSL;][]{Chen2014}. 
Although \citet{Mendigutia2014} refrained from using any line ratio to infer the MK class due to possible line veiling, the following reasoning led us to a slightly different conclusion.
For electron temperatures between 5000 K and 15000 K, the emission line region is expected to have a H$_{\delta}$:Br$_{\gamma}$ ratio of 5:1 \citep{Osterbrock2006}.
The NIR spectrum (Fig.~\ref{fig_spec_A}) gives a Br$_{\gamma}$ line intensity of at most $\sim 3.4 \times 10^{-10}$ ergs s$^{-1}$ cm$^{-2}$ $\mu$m. Therefore, the H$_{\delta}$ from the emission line region of the accretion
disk can be at most $\sim 1.7 \times 10^{-9} $ ergs s$^{-1}$ cm$^{-2}$ $\mu$m. The X-SHOOTER spectrum in the $H_{\delta}$ region indicates a continuum level of $\sim 3.2 \times 10^{-8}$ ergs s$^{-1}$ cm$^{-2}$ $\mu$m, which is thus more than 18 times brighter than the emission line expected from the disk region. 
Therefore, contrarily to the NIR Br$_{\gamma}$ line, the line veiling of H$_{\delta}$ will be negligible. 
Furthermore, the H$_{\delta}$ absorption line has about 50\% of the depth of the local continuum, which is in good agreement with all other XSL F stars with spectral type later than F4. 
In conclusion, following this line of thought, (i) the low level Paschen lines and ground state Ca II H+K are affected by line veiling due to the line strength; (ii) line veiling in Brackett lines cannot be neglected because the photospheric flux is much weaker at those wavelengths; and (iii) the Ca II IR triplet, the Balmer lines, and the higher level Paschen lines can be assumed to be marginally affected by the ongoing accretion. 

Therefore, we used the classical MK standard Fe I (4045\AA, 4226\AA)/H$_{\delta}$ ratio to estimate the spectral class of HD~142527~A.
Fe I (4045\AA, 4226\AA)/H$_{\delta}$ is $<$ 1 for F5 and earlier, and $\ge 1$ for F6 and later \citep{Morgan1978}. 
These lines are known to be fairly independent from the gravity/luminosity class. 
The measured line ratio for \object{HD~142527~A} is $\approx$ 1, which indicates a spectral type of F5-6. This is also the case for the XSL comparison stars, so possible instrumental and resolution effects can be excluded. 
Furthermore, the region of the Ca II IR triplet (8498\AA, 8542\AA, 8662\AA), the Paschen series line (Pa10, Pa11, Pa12, Pa13), and the Mg I (8807\AA) were used according to the observational calibrations of stars later than F2 by \citet{Marrese2003} and the synthetic spectra of \citet{Zwitter2004}.
The parametrization of the F star spectra from the XSL, including our target, was then executed as in 
\citet{Kordopatis2011}. This leads to a spectral type of F6-7.
Based on both criteria, we conclude that the spectral type must be F6 with an error of less than one subclass.

The luminosity class 
was estimated using two methods. First, we computed the classical MK system indicator Fe I (4063\AA) for our target and compared it to the F6V and F6III stars of the XSL. 
Second, we considered the depth of the higher level Paschen lines relative to the Ca II IR triplet, as it also depends strongly on gravity. %
Both methods indicate a gravity value significantly smaller than that of main sequence stars, but larger than those of giant stars. 
We therefore conclude that the luminosity class must be comprised between III and V, similarly to \citet{Mendigutia2014}. 

\section{Residual speckle uncertainty on the parameters of HD~142527~B} \label{app:uncertainties}

In order to evaluate the residual speckle uncertainty for the radial separation, PA and contrast of the companion derived with NEGFC, we first subtracted the companion from the original cubes using the optimal companion parameters found with NEGFC in each channel (Fig.~\ref{fig_npc-SNR}).
Then, for each channel, we injected one by one a series of 360 (positive) fake companions at the same median radial separation than
the companion, equally separated azimuthally by one degree. 
Each individual fake companion is injected at the inferred contrast of the companion, at a known radial separation and PA.
We then run again the NEGFC simplex algorithm with the 24 different sets of parameters detailed in Sect.~\ref{negfc}, spanning different aperture sizes, figures of merit and number of principal components used in PCA.
Next, we compute the median difference (over the 24 runs) between the injected companion parameters and the optimal parameters returned by NEGFC.
The 1$\sigma$--widths of the error histograms over the 360 fake companions can be considered as a good estimate of our residual speckle noise uncertainty \citep{Wertz2017}.

\begin{figure*}[tbh]
\begin{center}
\includegraphics[width=\textwidth]{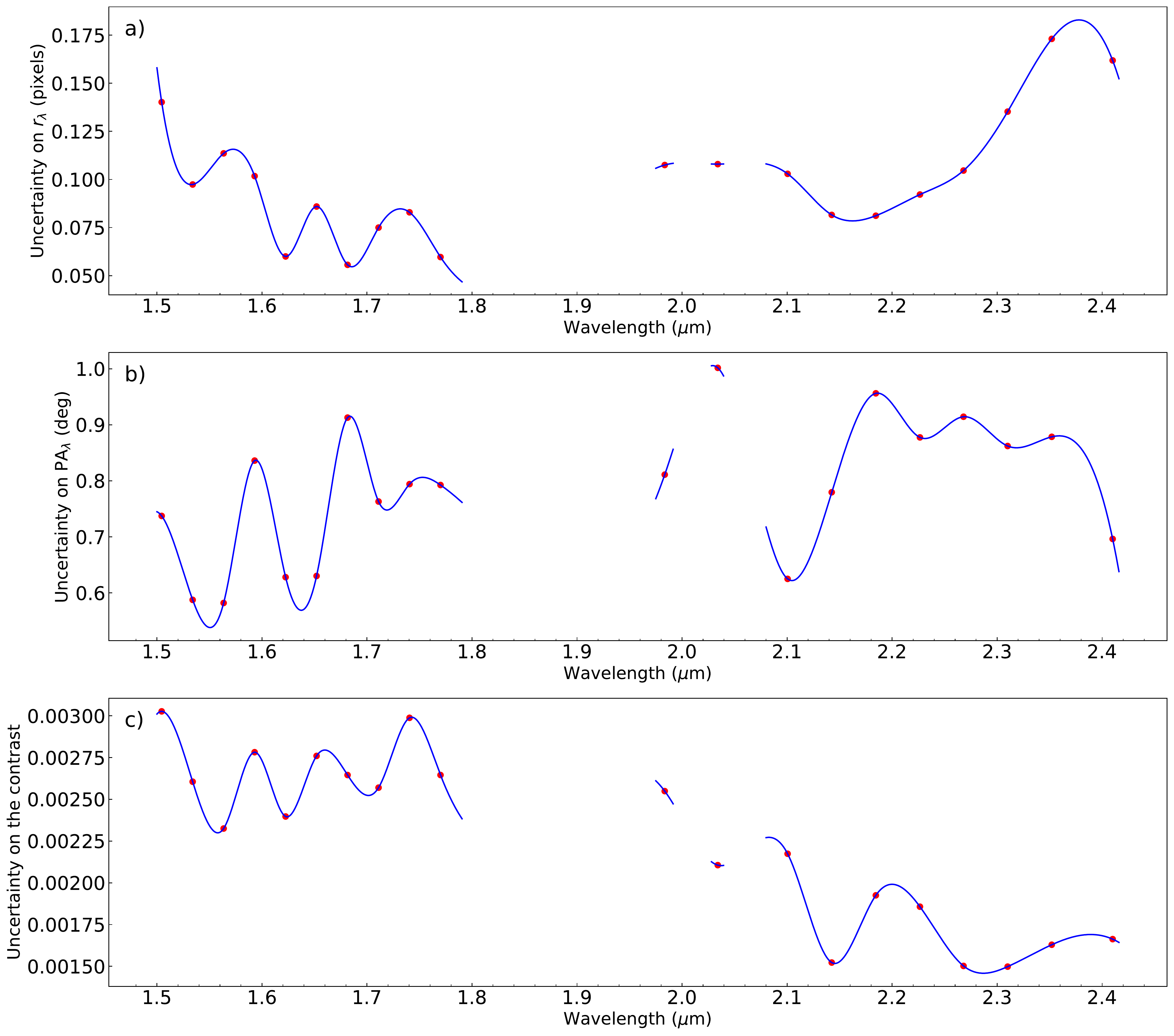} 
\end{center}
\caption{ \label{fig_speck_errors}
Estimated uncertainties on the parameters of HD~142527~B retrieved by NEGFC: {\bf a)} radial separation (in pixels),
{\bf b)} position angle (in degrees), and {\bf c)} contrast with respect to the primary in the different spectral channels. 
}
\end{figure*}

Given the large amount of resources required by the procedure described above, we only applied it to 20 different spectral channels spread over the H+K spectrum.
Considering the high correlation between neighboring spectral channels, the estimated uncertainties should be similar in neighboring channels.
Therefore, we interpolated the uncertainties to all spectral channels using a cubic spline.
The 1$\sigma$--uncertainties on the radial separation, PA and contrast of the companion are shown in Fig.~\ref{fig_speck_errors}a, b and c respectively.

\begin{figure*}[tbh]
\begin{center}
\includegraphics[width=\textwidth]{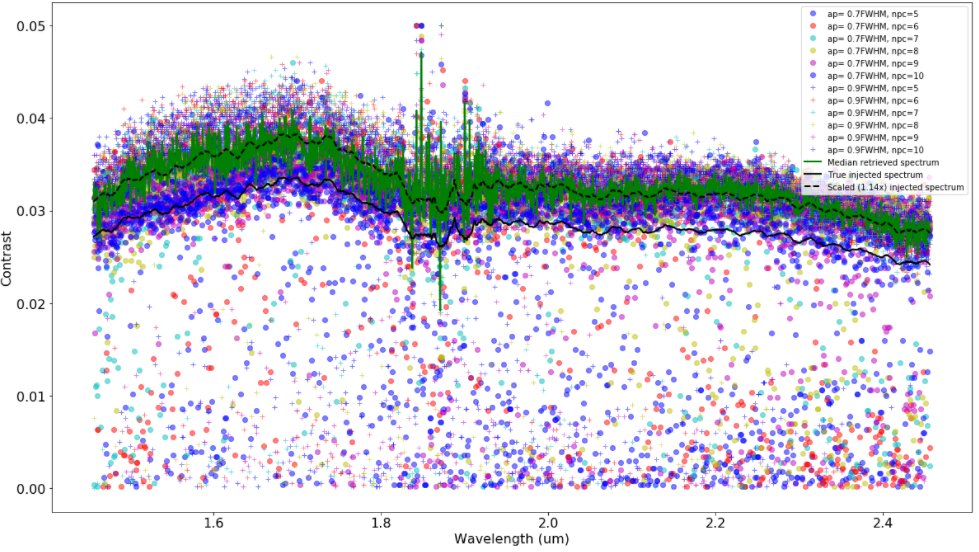} 
\end{center}
\caption{ \label{fig_speck_errors2}
Extracted spectrum of an artificially injected companion on top of the strongest speckle feature located at the same radial separation as the companion, at a PA of $240\degr$. 
The true injected spectrum is shown with the \emph{solid black curve}. 
The contrast of the companion is estimated in each spectral channel with different sets of NEGFC parameters: $n_{pc} \in [5,10]$ and apertures of 0.7 and 0.9 FWHM, using the \emph{standard deviation} as figure of merit (see details in Sect.~\ref{negfc}).
For clarity, we do not show the contrast estimates using the \emph{sum} as figure of merit, but do include these results in the median contrast spectrum obtained over all 24 reductions (\emph{green curve}).
The median retrieved contrast spectrum follows well the shape of the injected spectrum, but scaled up by a factor 1.14 (\emph{dashed black curve}).
}
\end{figure*}

A possible weakness in this method is that it assumes that the speckle pattern is randomly distributed in the image. 
Images in Fig.~\ref{fig_conspicuous}a to c might suggest a preferential triangular geometry for the speckle pattern, with the companion possibly located on top of one of the vertices.
Although this pattern is significantly damped when using PCA-annulus with sufficient principal components (Fig.~\ref{fig_conspicuous}d), we considered the worst case scenario and investigated the extraction of the spectrum of an artificially injected companion at the second brightest vertex of that triangle (the brightest after the companion itself).
We proceeded as follow. 
We first subtracted the companion from the original cubes using the optimal companion parameters found with NEGFC in each channel, then injected the artificial companion at the same radial separation but at a PA of 240$\degr$.
That artificial companion was injected at the same contrast as estimated by NEGFC for the real companion in each spectral channel (Fig.~\ref{fig_npc-SNR}c), but after smoothing it. 
The injected spectrum is shown by the \emph{black curve} in Fig.~\ref{fig_speck_errors2}.
We then used NEGFC with the 24 different sets of parameters to estimate the optimal radial separation, PA and contrast of the artificial companion in each spectral channel, in a similar way as for the real companion (see Sect.~\ref{negfc}).
For clarity, Fig.~\ref{fig_speck_errors2} only shows the 12 reductions corresponding to the \emph{standard deviation} figure of merit, $n_{pc} \in [5,10]$ and apertures of 0.7 and 0.9 FWHM. 
Analogously to the NEGFC reductions for the real companion, we notice the presence of a significant amount of outliers in individual channels, which justify the use of the median of the different runs (\emph{green curve}) to infer the contrast of the companion.
The shape of the median contrast spectrum is remarkably similar to the injected spectrum, as can be seen from the comparison to the injected spectrum scaled up by a factor 1.14 (\emph{dashed black curve}).
The speckle feature therefore only adds a global uncertainty on the flux of the companion of up to $15$\%, but does not affect significantly the shape of the extracted spectrum ($\sim 5$\% relative uncertainty).
We do not expect the uncertainty on the absolute flux of the true companion to be larger than $15$\% provided that injecting fake companions at all PA leads to 5--10 \% uncertainties on the contrast of all channels (Fig.~\ref{fig_speck_errors}c), that is, slightly more than 5\% but less than 15\%. 

Given that the 15\% uncertainty only affects the absolute flux level of the companion, but not the shape of the spectrum, we consider the 5 to 10 \% uncertainties mentioned in Fig.~\ref{fig_speck_errors}c for the best fit to BT-SETTL models (Sect.~\ref{BTSETTL}) and template spectra (Sect.~\ref{realspectra}).
However, we conservatively propagate the $15$\% flux uncertainty to the best fit radius of the companion, since it is the relevant parameter for scaling the BT-SETTL models (provided in units of flux at the stellar surface) to the observed flux of HD~142527~B.
In the case of the fit to the photosphere+hot environment model, the $15$\% uncertainty is included in the best-fit parameter ranges of $R_B$ and $R_d$ provided in Table \ref{tab:best_fit_params}.

\end{appendix}

\end{document}